\documentclass[aps, prb, twocolumn, superscriptaddress, english]{revtex4-2}

\usepackage{graphicx}
\usepackage{cancel}
\usepackage{enumerate}
\usepackage{hyperref}
\hypersetup{pdfencoding=auto,colorlinks=true,linkcolor=blue,citecolor=blue}
\usepackage{textcomp}
\usepackage{amsmath}
\usepackage{amssymb}
\usepackage{pgf}
\usepackage{soul}
\usepackage{subfigure}
\usepackage[english]{babel}
\usepackage{tikz}
\usepackage[normalem]{ulem}

\newcommand{\bbeta}{\boldsymbol{\beta}}
\newcommand{\bgamma}{\boldsymbol{\gamma}}
\newcommand{\bepsilon}{\boldsymbol{\epsilon}}

\def\ket#1{| #1\rangle}
\newcommand{\braket}[2]{\langle #1 | #2 \rangle}

\newcommand{\nep}{\textrm{e}}

\newcommand{\dQA}{\mathrm{\scriptscriptstyle dQA}}

\newcommand{\psitarget}{\psi_{\mathrm{\scriptscriptstyle targ}}}
\newcommand{\target}{\mathrm{\scriptscriptstyle targ}}
\newcommand{\mixing}{\mathrm{\scriptscriptstyle mix}}

\newcommand{\calA}{{\mathcal{A}}}

\newcommand{\calB}{{\mathcal{B}}}
\newcommand{\calC}{{\mathcal{C}}}

\newcommand{\Wilson}[1]{\mathcal{W}_{#1}}
\newcommand{\PauliSigma}{\hat{\sigma}}

\newcommand{\Ham}{\widehat{H}}

\newcommand{\Htarg}{\widehat{H}_{\scriptscriptstyle \mathrm{targ}}}

\newcommand{\Ptrot}{\mathrm{P}}

\newcommand{\Ztwo}{{\mathbb Z}_2}

\graphicspath{{./}{./FIgure/}}
\begin{document}

\title{Two-dimensional $\Ztwo$ lattice gauge theory on a near-term quantum simulator: variational quantum optimization, confinement, and topological order.}

\author{Luca Lumia}
\affiliation{SISSA, Via Bonomea 265, I-34135 Trieste, Italy}
\author{Pietro Torta}
\affiliation{SISSA, Via Bonomea 265, I-34135 Trieste, Italy}
\author{Glen B. Mbeng}
\affiliation{Universit\"at Innsbruck, Technikerstra{\ss}e 21 a, A-6020 Innsbruck, Austria}
\author{Giuseppe E. Santoro}
\affiliation{SISSA, Via Bonomea 265, I-34135 Trieste, Italy}
\affiliation{International Centre for Theoretical Physics (ICTP), P.O.Box 586, I-34014 Trieste, Italy}
\affiliation{CNR-IOM Democritos National Simulation Center, Via Bonomea 265, I-34136 Trieste, Italy}
\author{Elisa Ercolessi}
\affiliation{Dipartimento di Fisica, Universit\`a di Bologna and INFN, Via Irnerio 46, 40126 Bologna, Italy}
\author{Michele Burrello}
\affiliation{Niels Bohr International Academy and Center for Quantum Devices, Niels Bohr Institute, Copenhagen University, Universitetsparken 5, 2100 Copenhagen, Denmark}
\author{Matteo M. Wauters}
\affiliation{Niels Bohr International Academy and Center for Quantum Devices, Niels Bohr Institute, Copenhagen University, Universitetsparken 5, 2100 Copenhagen, Denmark}

\begin{abstract}    
We propose an implementation of a two-dimensional $\Ztwo$ lattice gauge theory model on a shallow quantum circuit, involving a number of single and two-qubits gates comparable to what can be achieved with present-day and near-future technologies. 
    The ground state preparation is numerically analyzed on a small lattice with a variational quantum algorithm, which requires a small number of parameters to reach high fidelities and can be efficiently scaled up on larger systems.
    Despite the reduced size of the lattice we consider, a transition between confined and deconfined regimes can be detected by measuring  expectation values of Wilson loop operators or the topological entropy. 
    Moreover, if periodic boundary conditions are implemented, the same optimal solution is transferable among all four different topological sectors, without any need for further optimization on the variational parameters. 
    Our work shows that variational quantum algorithms provide a useful technique to be added in the growing toolbox for digital simulations of lattice gauge theories.
\end{abstract}
\maketitle
\section{Introduction}
Platforms for quantum computation are undergoing steady development and, nowadays, several architectures for the implementation of quantum circuits up to a few tens of qubits are available. In most of these platforms, qubits are considerably noisy, with coherence times that enable the application of only a few layers of unitary gates before losing quantum correlations. Therefore, for a successful use of these quantum computers the task of efficiently initializing the system in a target state becomes of the uttermost importance.

Variational quantum algorithms (VQAs)~\cite{Wecker_PRA2015,Cerezo_NatRev2021} are a promising technique to exploit noisy quantum hardware with shallow circuits.
Among such methods, the quantum approximate optimization algorithm~\cite{Farhi_arXiv2014,Mbeng_arx19,Zhou_PRX2020} (QAOA) is a popular strategy for preparing the ground state of a target many-body Hamiltonian $\Ham_{\target}$. 
Starting from a product state, usually the ground state of a simple mixing Hamiltonian $\Ham_{\mixing}$, it applies a series of unitary evolution operators generated alternately by $\Ham_{\target}$ and $\Ham_{\mixing}$. 
The corresponding evolution times are treated as variational parameters to be optimized via a classical minimization of the energy.   
QAOA has been studied extensively for solving classical optimization problems~\cite{Farhi_arXiv2014,Zhou_PRX2020,Zhu_arxiv2020}, ground state approximation of quantum spin systems~\cite{Mbeng_arx19,Ho_SciPost2019,Ho_PRA19,Wauters_PRA2020,Matos_PRXQ2021}, and has been shown to be computationally universal
~\cite{Lloyd_arXiv2018,Morales_QInfoProc2020}.
Moreover, it has been successfully implemented on a trapped-ions quantum simulator~\cite{Pagano_PNAS2020} and on superconducting quantum circuits~\cite{Alam_arxiv2019,Karamlou_NPJ2021}.

When considering current quantum computers based on superconducting qubits, the typical platforms are constituted by qubits arranged in two-dimensional arrays~\cite{Arute_Nature2019,Jurcevic2021}. These systems allow us to manipulate the qubits via single and two-qubit gates and, in most cases, two-qubit gates are local, i.e. they can be applied on neighbouring qubits only. These platforms are usually initialized starting from simple product states, while the efficient creation of complex entangled states is a non-trivial challenge. 
Despite the limitations of the noisy hardware currently available, these architectures open the path to
various applications. One of the most important is the implementation of error-correcting codes and, to this purpose, considerable attention has been devoted to the realization of surface codes \cite{Fowler2012}, thus to the preparation via local gates of states displaying topological order. In this context, the first experimental realization of the ground state of a surface code has been successfully achieved \cite{Satzinger_Science2021} and it allowed for the study of the main topological properties of its anyonic excitations.
Another strategic application is the simulation of interacting quantum many-body problems that are intrinsically hard to simulate with classical computers~\cite{Lloyd_qsimulators}.

In this framework, lattice gauge theories (LGT) emerged as a paradigmatic research subject. They constitute the backbone of particle physics, and many of their important features display a non-perturbative nature, requiring therefore advanced numerical techniques to be studied. Furthermore, some of the simplest two-dimensional (2D) LGTs share the same topological properties of surface codes which, indeed, can be seen as the extreme deconfined limit of systems with $\Ztwo$ gauge symmetry. In the last decade, the application of quantum technologies to LGT became a lively field of research \cite{zohar2015,dalmonte2016,banuls2020,banuls2020b,Zohar_PhilTransA2021,Aidelsburger_LGT2021}, progressing both on the development of several technologies and algorithms to tackle the complexity of LGTs and on the study of LGTs themselves.

One of the tasks which can be addressed through quantum simulation of LGTs is the study of their static properties.
To this purpose, a key step is an efficient initialization of their ground states, allowing the investigation of their phase diagrams at low temperatures.

In this work, we explore the possibility of studying the ground state of a 2D pure lattice gauge theory within the framework of quantum circuits and digital quantum simulations.
Indeed, the recent developments of quantum simulations provide complementary approaches to other quantum many-body approximation techniques such as Tensor Networks~\cite{Tagliacozzo_PRX2014, Silvi2014, banuls2020, Celi_PRX2020, Felser_PRX2020, Zohar_PRR2021}, which are challenging to implement in 2D with current technologies, in particular for what regards quantum dynamics.
We shall focus on the 2D $\Ztwo$ LGT, which is known to display a confinement-deconfinement phase transition between a trivial (confined) phase and a topologically ordered (deconfined) phase, matching the topological features of surface codes. 
As we show, most of the interesting ground state properties linked to topological order, which are usually described in the thermodynamic limit, can be characterized even with small lattices.

Our approach to realize the ground states of a 2D LGT is based on single and two-qubit gates only, and it is amenable to experimental implementations with limited resources, in particular with a number of qubits and fan-out already accessible to present-day hardware or near-future devices.
We apply QAOA to prepare the ground state at arbitrary values of the coupling and we show that the algorithm reaches high fidelities within a very small number of variational parameters, corresponding to a shallow quantum circuit. 
To reliably find optimal or quasi-optimal minima, we employ a two-step local optimization procedure~\cite{QAOA_perceptron}, which provides regular schedules that can be efficiently transferred to larger systems.

Targeting the ground state in the confined phase, where there is no long-range entanglement, can always be performed efficiently and our numerical simulations suggest that QAOA can be scaled up to larger sizes without increasing the circuit depth.
Concerning the preparation of states in the deconfined phase, instead, it is known that topologically ordered states cannot be obtained exactly with circuits of fixed depth for growing system size. In particular, for the ground states of the toric code, the required circuit depth scales linearly with the system width \cite{Bravyi_PRL2006,Chen2010,Bravyi_PRL2019,Satzinger_Science2021,Liu_arxiv2021}.
This is also a general property of QAOA, where long-range correlation and perfect control on the system is attained only with an extensive number of layers~\cite{Mbeng_arx19, Farhi_arxiv2020_1, Farhi_arxiv2020_2}.  

As a consequence, when targeting states in different phases, we compare two strategies: either we apply directly the QAOA evolution on a trivial product state or we first build exactly the toric code state and then apply the variational circuit from that starting point.
The two approaches offer optimal results for targeting states in the confined or deconfined phase, respectively. They also display similar performances for the small system sizes we consider, except for the required overhead of the second approach.

The ground states prepared with QAOA are then used to characterize the crossover from the confined-topologically trivial phase to the deconfined-topologically ordered one.
In particular, we focus on the behavior of Wilson loop operators and of the topological entropy. We also discuss the possibility of exploring the ground state degeneracy when the lattice has periodic boundary conditions.
Remarkably, all these indicators of a topological phase transition display very small deviations from their expected behavior in the thermodynamic limit despite the reduced size of our lattices.
The successful implementation of 2D $\Ztwo$ LGT and the correct description of this non-trivial crossover --- with very limited resources in terms of qubit numbers and circuit depth --- provides a proof of principle of the feasibility of quantum simulations of deconfined and topological phases of lattice gauge theories in general.  

The rest of this article is organized as follows.
In Sec.~\ref{sec:model} we describe the $\Ztwo$ lattice gauge theory model, introducing the main properties that characterize its topological order.
In Sec.~\ref{sec:alg} we describe the implementation of QAOA for this model and how to represent the relevant unitary operators in terms of single and two-qubit gates.
The main numerical analysis is presented in Sec.~\ref{sec:results}, where we focus on the performance of QAOA, its scalability to larger systems, and the characterization of the topologically ordered phase.
Finally, we summarize our main results and some future perspectives in Sec.~\ref{sec:conclusion}.

\section{$\Ztwo$ lattice gauge theory}\label{sec:model}

\begin{figure}
	\centering
\includegraphics[width=6cm]{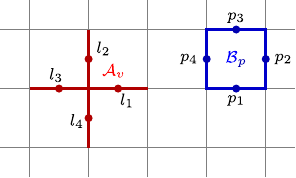}
 \caption{Representation of a star operator $\calA_v$ (in red) and a plaquette operator $\calB_p$ (in blue), with the corresponding qubits on the links (solid circles).
 }
 \label{fig:star-plaquette}
\end{figure}

In this work, we consider a pure $\Ztwo$ gauge theory model on a regular square lattice. The discretized gauge fields are represented by qubits on the links of the lattice.
Using a lattice of size $L\times L$, there are $2L^2$ qubits if periodic boundary conditions are imposed.
The Hamiltonian we use is the sum of two competing terms
\begin{equation}\label{eq:Htot}
    \Ham = \Ham_E + h \,\Ham_B \ ,
\end{equation}
which represent “electric” and “magnetic” non-commuting contributions. Their structure comes from an analogy with the QED Hamiltonian, where both space and the gauge group $U(1)$ are discretized: the real space becomes a lattice, and $U(1)$ is discretized to $\mathbb{Z}_n$. Here we focus on the smallest discrete group $n=2$, which is naturally encoded in terms of qubits.
The electric contribution to the Hamiltonian is 
\begin{equation}\label{eq:HE}
    \Ham_E = \sum_l(1 - \PauliSigma^x_l) \ , 
\end{equation}
where the index $l$ runs over all the links in the lattice and the Pauli matrices are denoted $\PauliSigma^\alpha_{l}$, with $\alpha=x,\, y,\, z$. 
This specific choice of $\Ham_E$ is motivated by the QED analogy, since the electric field enters the Hamiltonian via $\vec{E}^2$ and our term is positive definite. 
A spin in the eigenstate  $\PauliSigma^x_l\ket{+}_l = \ket{+}_l$ brings no contribution to the electric energy and corresponds to a vanishing electric field. 
The state $\ket{-}_l$ indicates instead the presence of a $\Ztwo$ electric excitation on the link $l$, with energy cost assigned by $\Ham_E$.
The magnetic term reads
\begin{equation}\label{eq:HB}
    \Ham_B = -\sum_{p} \calB_p = -\sum_{p} \PauliSigma^z_{p_1} \PauliSigma^z_{p_2} \PauliSigma^z_{p_3} \PauliSigma^z_{p_4} \ ,
\end{equation}
where $p$ labels the {\em plaquettes} of the lattice and the \emph{plaquette operator} $\mathcal{B}_p$ involves the product of the four spin variables $\PauliSigma^z$ around the four links $p_1,\cdots p_4$ of the $p$-th plaquette (see Fig. \ref{fig:star-plaquette}). 
In particular, $\mathcal{B}_p= -1$ represents a magnetic flux through the $p$-th plaquette, and the interaction term $h\Ham_B$ assigns an energy $2h$ to each of these excitations. 
The electric Hamiltonian $\Ham_E$ effectively provides a kinetic energy to the magnetic fluxes.
The local gauge constraint is the analog of Gauss's law and it selects the physically relevant sector of the Hilbert space.
For each vertex $v$ of the lattice, physical states must be left invariant by gauge transformations, thus satisfying
\begin{equation}\label{eq:gauge}
    \calA_v \ket{\psi}_{\rm phys} = \prod_{l \in v} \PauliSigma^x_l \ket{\psi}_{\rm phys}=\ket{\psi}_{\rm phys} \ ,
\end{equation}
where the \emph{star operator} $\calA_v$ is the product of the spin operators $\PauliSigma^x$ on the four links  connected to the vertex $v$, as represented in Fig. \ref{fig:star-plaquette}.

The Hamiltonian in Eq.~\eqref{eq:Htot} has two well known limits, for $h\to 0$ and $h\to \infty$.
When only $H_E$ is present ($h\to 0$), the electric ground state is a trivial product state with all spins aligned along the $x$ direction $\ket{\Omega_E} = \bigotimes_l \ket{+}_l$, which satisfies the local gauge constraints in Eq.~\eqref{eq:gauge} and corresponds to the absence of any electric field excitation.
In the opposite limit $h \to \infty$ only the magnetic term remains, the system behaves like a surface code \cite{kitaev2003,Bravyi1998,Fowler2012} and displays topological order. In this case, the number of ground states depends on the boundary conditions. The ground states of $\Ham_B$ are the simultaneous eigenstates of all plaquette and star operators with eigenvalue 1 and correspond to the absence of magnetic fluxes.

In the case of open and smooth  boundaries \cite{Fowler2012}, there is a single magnetic ground state which can be expressed as an equal amplitude linear superposition of all possible contractible electric flux loops $\Gamma$:
\begin{equation} \label{eq:Omega_B}
    \ket{\Omega_B} = \mathcal{N} \, \sum_\Gamma \mathcal{W}_\Gamma \ket{\Omega_E} = \prod_p  \left( \frac{1 + \mathcal{B}_p}{\sqrt{2}}\right) \ket{\Omega_E}\ .
\end{equation}
%
Here $\mathcal{N}$ is a normalization factor and $\mathcal{W}_\Gamma$ the Wilson loop operator associated to a closed path $\Gamma$,  defined as the product of $\PauliSigma^z$ matrices on the links belonging to $\Gamma$:
\begin{equation} \label{eq:Wilson}
    \mathcal{W}_\Gamma  =\prod_{l\in \Gamma} \,\PauliSigma^z_l \ .  
\end{equation}
%
Since $\PauliSigma^z\ket{\pm}=\ket{\mp}$, Wilson loops applied to the electric ground state $\ket{\Omega_E}$ create closed lines of electric field excitations. 
The second form of $\ket{\Omega_B}$ in Eq.~\eqref{eq:Omega_B} expresses it as the normalized product of the projectors on the eigenstate of each plaquette operator with eigenvalue 1. 

The magnetic coupling $h$ drives the system across a topological phase transition, occurring at $h_c$, between the electric and the magnetic phases, which are distinguished by different behaviors of the expectation values of the Wilson loop operators. 
From the definitions of the limiting ground states $\ket{\Omega}_E$, $\ket{\Omega}_B$, it follows that in the two limits $h=0$ and $h\to \infty$ we have for all paths $\Gamma$
%
\begin{equation*}
\braket{\Omega_E}{\mathcal{W}_\Gamma|\Omega_E}=0\hspace{5mm} \mbox{and} \hspace{5mm}
\braket{\Omega_B}{\mathcal{W}_\Gamma|\Omega_B}=1 \;.
\end{equation*}
%
At a finite value of $h$, the expectation value of Wilson loops on the ground state decreases exponentially in the size of $\Gamma$, with a leading contribution given by~\cite{fradkin1978, Greensite2003}
\begin{equation}\label{eq:wilson_gen}
    \langle \Wilson{\Gamma} \rangle = \nep^{-\chi(h) A_\Gamma-\delta(h) P_\Gamma} \ ,
\end{equation}
where $A_\Gamma$ and $P_\Gamma$ are the area enclosed by the loop $\Gamma$ and its perimeter, respectively, while $\chi$ and $\delta$ are two positive functions. For $h<h_c$, the system is in a phase dominated by the electric term $\Ham_E$, $\chi(h)>0$ and $\langle \Wilson{\Gamma} \rangle$ decays with an ``area law''.
This means that large loops of electric excitations are strongly suppressed, which is a signature of \emph{confinement} \cite{kogut1979,Creutz_PRL1980}.
In the opposite {\em deconfined} phase, where $h>h_c$ and the dominant term is $\Ham_B$, $\chi(h) \to 0$, and the behavior of large Wilson loops follows a ``perimeter law''.
 
\begin{figure}
	\centering
	\includegraphics[width=8cm]{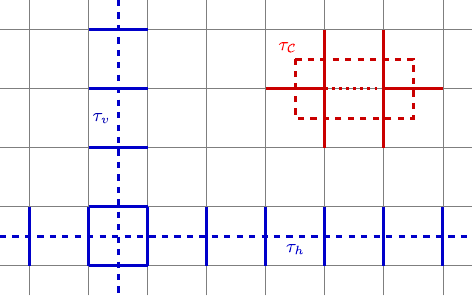}
	\caption{The two non-contractible 't Hooft loops $\tau_h,\,\tau_v$ and a simple example of how a contractible 't Hooft loop operator $\tau_{\calC}$ is written as a product of star operators. The dotted link indicates a cancellation due to $(\PauliSigma^x_l)^2=1$.}
\label{fig:thooft}
\end{figure}

When periodic boundary conditions (PBC) are considered in both directions, the Hamiltonian acquires an extra $\Ztwo \times \Ztwo$ symmetry related to non-contractible 't Hooft loops.
Consider a closed path $\calC$ in the dual lattice. $\calC$ crosses orthogonally a sequence of links of the direct lattice that we denote schematically by $\ell\cap\,\calC$. 
The 't Hooft loop operator
\begin{equation}
    \tau_\calC = \prod_{l\, \cap\,\calC} \,\PauliSigma^x_l\,,
\end{equation}
commutes with the Hamiltonian \eqref{eq:Htot} for any closed loop $\calC$. However, if $\calC$ is contractible, $\tau_\calC$ can always be expressed as a product of star operators $A_v$, so that $[\tau_\calC,\hat{H}]=0$ does not provide any additional information that is not already contained in the gauge-invariance of the Hamiltonian. 
Considering PBC, the lattice becomes a torus and there are indeed two inequivalent non-contractible loops, whose corresponding 't Hooft operators $\tau_h$ and $\tau_v$ provide new symmetries. Fig.~\ref{fig:thooft} shows examples of 't Hooft loop operators.

In the trivial limit $h\to 0$, the expectation values of $\tau_h$ and $\tau_v$ on the electric ground state $\ket{\Omega_E}$ are $+1$. Switching on strings of electric excitations, one sees that $\tau_h$ and $\tau_v$ measure the winding number of the electric field lines in the vertical and horizontal directions, respectively.
With PBC, we can construct two non-contractible Wilson loop operators $\Wilson{h}$ and $\Wilson{v}$, which commute with $\Ham_B$ and $\calA_v$, but vary the electric field winding numbers since they respectively anticommute with $\tau_v$ and $\tau_h$: 
\begin{eqnarray}
[\Wilson{h},\tau_h] &=& [\Wilson{v},\tau_v] = 0 \nonumber \\
\{\Wilson{h},\tau_v\} &=& \{\Wilson{v},\tau_h\} = 0 \,.  
\end{eqnarray}
In the topological limit $h\to \infty$, when  $[\Wilson{h},\hat{H}]=[\Wilson{v},\hat{H}]=0$,
we get four degenerate ground states characterized by different eigenvalues of the 't Hooft loops, corresponding in the basis $\ket{\tau_h,\tau_v}$ to:
\begin{align*}
    \ket{++} &= \ket{\Omega_B} \ ,\\
    \ket{+-} &= \Wilson{h}\ket{\Omega_B}  \ ,\\
    \ket{-+} &= \Wilson{v}\ket{\Omega_B}\ ,\\
    \ket{--} &= \Wilson{h}\Wilson{v}\ket{\Omega_B}\ .
\end{align*}
At finite values of $h>h_c$, the perfect degeneracy between these four states gets lifted by an energy splitting vanishing exponentially with $L$.

The phase transition occurs at $h_c=3.04438(2)$ \cite{Blote_PRE2002} and it can be understood by considering the duality between the Hamiltonian in Eq.~\eqref{eq:Htot} and the 2-dimensional quantum Ising model with transverse field (2D-TFIM), whenever the lattice has PBC, the gauge-symmetry $A_v|\psi\rangle_{\rm phys}=|\psi\rangle_{\rm phys}$ is imposed at each vertex, and the Hilbert space is restricted to the $\tau_h=\tau_v=+1$ sector~\cite{Fradkin_book}.
Indeed, we can define new Pauli spin variables ${\rm X}_p $ and ${\rm Z}_p$ on the dual lattice, where
$p$ denotes the plaquette centers, 
by identifying 
\begin{equation} \label{eq:mapping}
    \left\lbrace
    \begin{array}{rcl}
        {\rm X}_p  &=& \calB_p \vspace{2mm} \\
        {\rm Z}_{p}{\rm Z}_{p'}  &=& \PauliSigma^x_{l(p,p')}
    \end{array}
    \right. \;,
\end{equation}
where $p$ and $p'$ are neighboring plaquettes and  $l(p,p')$ is the link shared by $p,\,p'$.
With the mapping in Eq.~\eqref{eq:mapping}, the Hamiltonian becomes a transverse-field Ising model on the dual lattice:
\begin{equation}\label{eq:Hising}
    \Ham = \sum_{\langle p, p' \rangle}\big(1- {\rm Z}_{p}{\rm Z}_{p'}\big) - h \sum_p  {\rm X}_p \;.
\end{equation}
%
One can check that the algebra generated by the new operators is the same as the original one, confirming the unitary equivalence of the two models.
Notice that the number of degrees of freedom is now halved: $L^2$ qubits (one for each plaquette), instead of $2L^2$ (one for each link). This is an effect of the gauge symmetries, which are now automatically incorporated in the model.
Finally, it is important to mention the fact that this duality fixes the global $\Ztwo$ symmetry of the Ising model: in the original representation, the product of all plaquette operators is the identity $\prod_p \calB_p = 1$.
In the 2D-TFIM, this is reflected in the condition $\prod_p \PauliSigma^x_p \ket{\psi}=\ket{\psi}$, which means that the physical states must be invariant under a global spin flip.

In this work, we shall employ both the original formulation of Eq.~\eqref{eq:Htot} and its dual model in Eq.~\eqref{eq:Hising}. The dual Ising model will be exploited to speed up our numerical analysis of the $\Ztwo$ LGT and, in particular, to verify the scalability of QAOA between different system sizes. 

\section{QAOA and circuit implementation}\label{sec:alg}
\subsection{Ground state preparation with QAOA}
To prepare the ground state of the LGT Hamiltonian in Eq.~\eqref{eq:Htot}, we use the Quantum Approximate Optimization Algorithm (QAOA)~\cite{Farhi_arXiv2014}.
Although QAOA was initially proposed as a tool to look for approximate solutions to classical combinatorial optimization problems, it can be easily generalized to construct the ground state of many-body quantum Hamiltonians.
Considering the two terms $\Ham_B$ and $\Ham_E$ in the LGT Hamiltonian, QAOA consists in constructing the following variational {\em Ansatz} 
\begin{equation} \label{eq:QAOA_state}
\ket{\psi_{\Ptrot}(\bgamma,\bbeta)} 
=  \widehat{U}(\gamma_\Ptrot,\beta_\Ptrot) \dots  \widehat{U}(\gamma_1,\beta_1)\, \ket{\psi_0} \;,
\end{equation}
where $\bbeta=\beta_1,\dots,\beta_{\Ptrot}$ and $\bgamma =\gamma_1,\dots,\gamma_{\Ptrot}$ are $2\Ptrot$ free real parameters, and the unitary operators $\widehat{U}(\gamma_{m},\beta_{m})$, for $m=1\cdots \Ptrot$, evolve the state according to $\Ham_{B}$ and $\Ham_{E}$, in an alternating fashion.
More precisely, the initial state $\ket{\psi_0}$ can either be the electric ground state $\ket{\Omega_E}$ or the magnetic one $\ket{\Omega_B}=\ket{++}$, and,
depending on the choice of $\ket{\psi_0}$, we define the operator $\widehat{U}_m=\widehat{U}(\gamma_{m},\beta_{m})$ in Eq.~\eqref{eq:QAOA_state} as
\footnote{This choice avoids adding only a global phase in the first step.}
\begin{equation}\label{eq:U_m}
     \widehat{U}_m =\left\{
    \begin{split}
        & \nep^{-i \beta_{m} \Ham_E } \nep^{-i \gamma_{m} \Ham_B} \hspace{3mm} {\rm if} \hspace{3mm} \ket{\psi_0}=\ket{\Omega_E} \,, \\
        & \nep^{-i \gamma_{m} \Ham_B} \nep^{-i \beta_{m} \Ham_E } \hspace{3mm} {\rm if} \hspace{3mm} \ket{\psi_0}=\ket{\Omega_B} \;. \\
    \end{split}
    \right.
\end{equation}
For a given choice of the coupling $h$, which identifies a target Hamiltonian $\Htarg(h) = \Ham_E + h \Ham_B$, an approximation of the associated ground state is found using a classical minimization of the variational energy
\begin{equation} \label{eq:QAOA_en}
    E_\Ptrot \left(\bgamma,\bbeta \right) = \langle \psi_{\Ptrot}(\bgamma,\bbeta) \,|\, \Htarg(h)\, |\, \psi_{\Ptrot}(\bgamma,\bbeta) \rangle
\end{equation}
in this 
$2\Ptrot$-dimensional energy landscape.

The optimal energy at the global minimum $E_\Ptrot \left(\bgamma^*,\bbeta^* \right)$ is a monotonically decreasing function of $\Ptrot$.
However, determining exactly the global minimum is, in general, not a trivial task~\cite{Zhou_PRX2020}, 
since local optimization routines tend to get trapped into one of the many local minima of the $2\Ptrot$-dimensional search space.
We will discuss below an effective strategy to search for optimal (or quasi-optimal) solutions by a two-step QAOA procedure that starts from a linear schedule for the parameters, in the spirit of a digitized Quantum Annealing~\cite{Mbeng_PRB2019}.

\subsection{Circuit implementation of the QAOA {\em Ansatz}}\label{ssec:circuit}
The QAOA variational wavefunction in Eq.~\eqref{eq:QAOA_state} is obtained by applying $\Ptrot$ layers of local unitary operators, by alternating the time evolutions generated by plaquette and electric field interactions.
In what follows, we are going to describe how to implement the operations involved in each layer of the variational circuit by using only single and two-qubit gates.
Since we shall focus on a single application of the unitary operations $\nep^{-i\beta_m\Ham_E}$ and $\nep^{-i\gamma_m\Ham_B}$, the index $m$ will be dropped from the parameters.

\begin{figure}
    \centering
    \includegraphics[width=8.5cm]{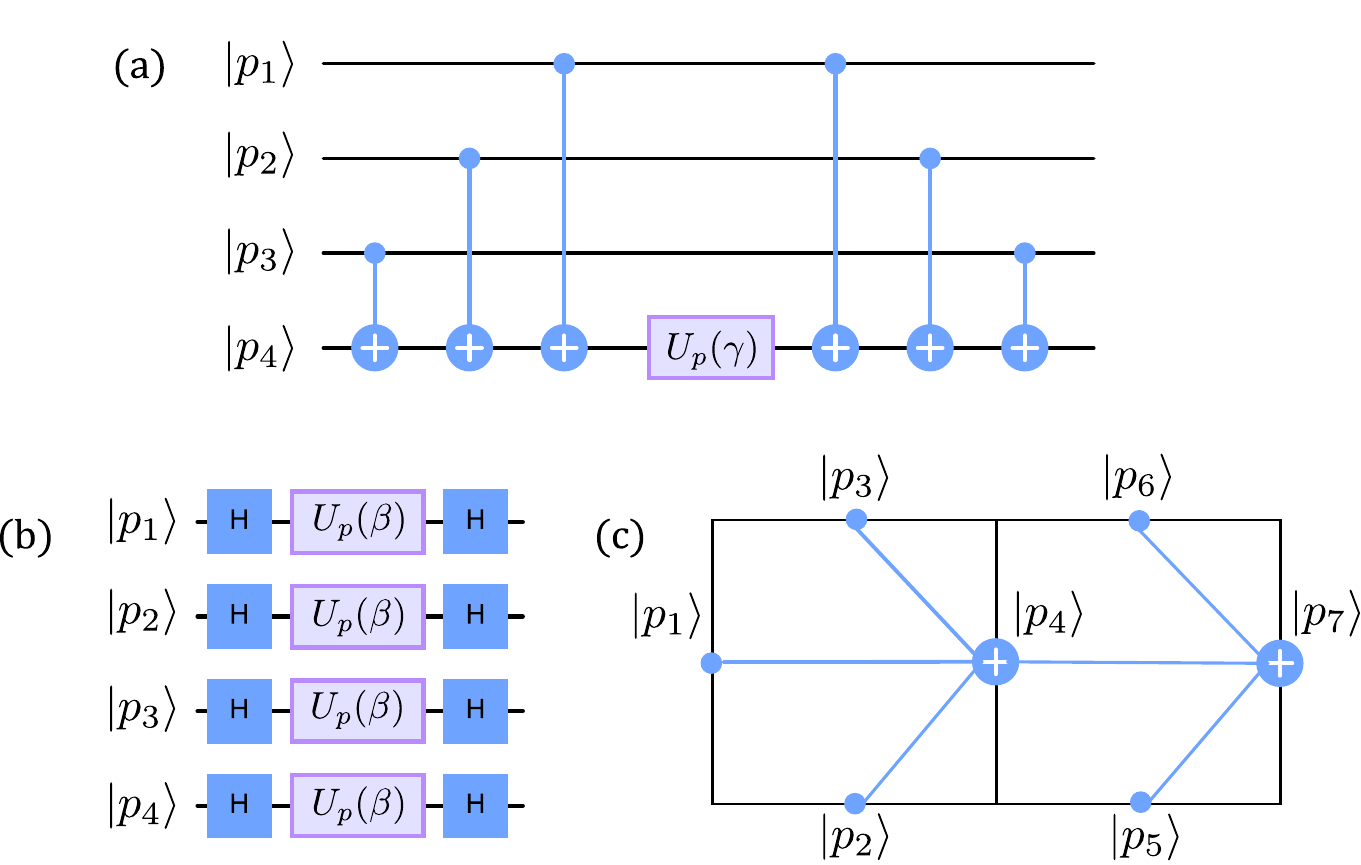}
    \caption{(a): Circuit implementation of the operator $\nep^{-i\gamma \Ham_B}$ acting on a single plaquette, with a target and 3 control qubits. The states $\ket{p_i}$ are expressed in the $\hat{\sigma}_z$ eigenbasis and $U_{p}(\gamma)=\nep^{ i\gamma \PauliSigma^z }$ is a single-qubit rotation around the $z$ axis.
    (b): implementation of the single-qubit operations that describe $\nep^{-i\beta\Ham_E}$. 
    (c): example of the {\em sequential} implementation of the operator $\nep^{-i \gamma \Ham_B}$ on two neighboring plaquettes. The qubit $\ket{p_4}$ is used first as the target qubit for the first plaquette and aftwerwards as a control qubit for the second.}
    \label{fig:circuit_sketch}
\end{figure}

The electric term of Eq. \eqref{eq:HE} is a sum of single-qubit operators, therefore the evolution it generates can be realized as a product of single-qubit rotations around the $x$ axis by the angle $\beta$, up to an irrelevant global phase. 
The computational basis we adopt hereafter is the $\PauliSigma^z$ eigenbasis. Therefore, we employ Hadamard gates to diagonalize $\Ham_E$, and we reproduce the electric evolution during a single QAOA step by simultaneously applying operators $U_p(\beta)=\nep^{i\beta \PauliSigma^z}$ to all qubits, i.e. a global rotation of angle $\beta$ around the $z$-axis~\footnote{In the $\PauliSigma^z$ eigenbasis the circuit  depth is 3; however, in a realistic experimental setup, it is often possible to implement single-qubit rotations around the $x$ axis, requiring only a single layer of gates}.
A schematic representation of the single-qubit gates required is sketched in Fig.~\ref{fig:circuit_sketch}(b).

The implementation of the time evolution associated with plaquette operators is less trivial \cite{Dlaska_PRL2021,Lechner_IEEE2020,lamm}, but it can be realized in a local way as a combination of single- and two-qubit gates. Fig.~\ref{fig:circuit_sketch}(a) shows that a single-plaquette unitary operator $\nep^{i\gamma \calB_p}$ is obtained by a suitable combination of CNOT gates and a single-qubit rotation $U_{p}(\gamma)$
applied to the fourth qubit of the plaquette. 
The fourth qubit is the target of all CNOTs and it is restored to its initial logical state by the last three CNOTs, such that $U_{p}(\gamma)$ successfully applies the phase $\gamma\calB_p$ only. An alternative technique based on ancillary qubits is presented in Refs. \cite{Zohar_PRL2017,Zohar_PRA2017,Armon_PRL2021}.

For a lattice composed of several plaquettes, the circuits in Fig. \ref{fig:circuit_sketch}(a) cannot be simultaneously run on all of them, since two neighboring plaquettes share a qubit: as shown in Fig.~\ref{fig:circuit_sketch}(c), the qubit 4 not only acts as the target for the left plaquette but also as one of the controls for the right plaquette.
The time evolution of the plaquette operators, however, can still be efficiently parallelized.
For the sake of simplicity we shall consider first systems with an even number of columns (for an even number of rows the situation 
is formally equivalent).
In this case, we can decompose the whole lattice into sets of two neighboring horizontal plaquettes, each set with the same structure as depicted in Fig.~\ref{fig:circuit_sketch}(c).

\begin{figure}
    \centering
    \includegraphics[width=8.5cm]{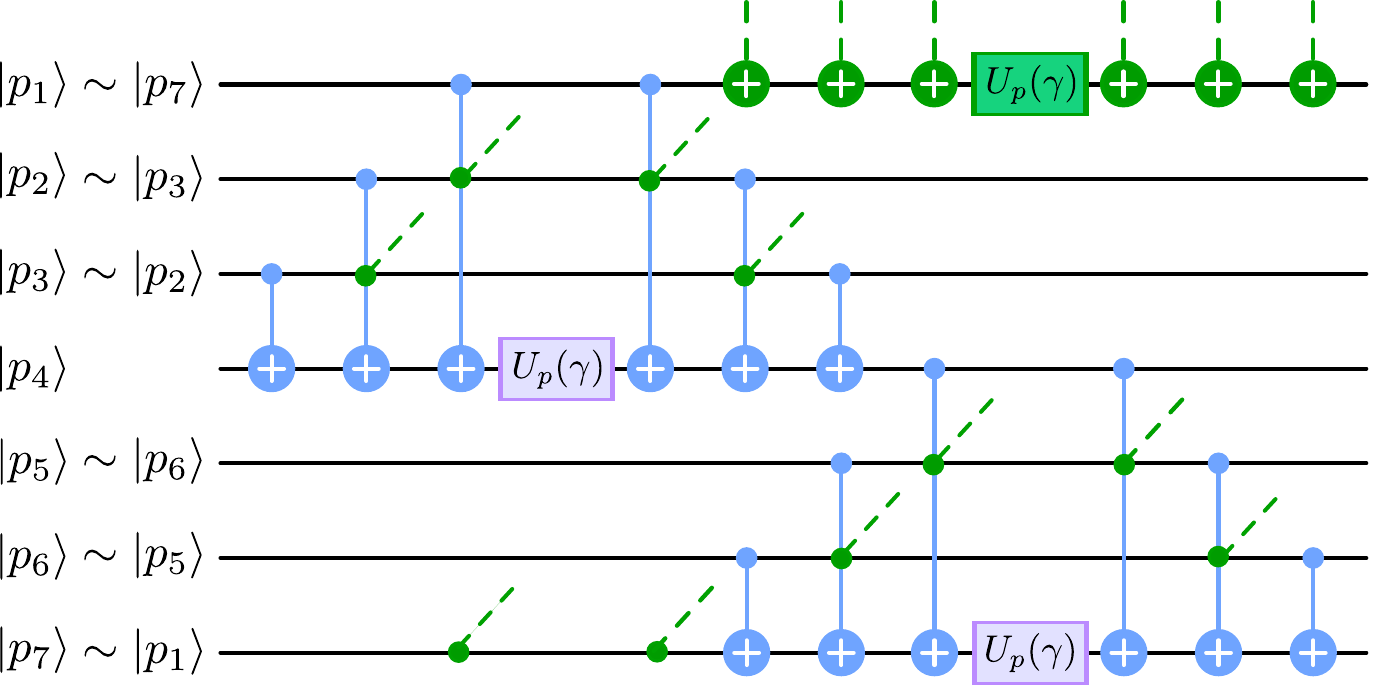}
    \caption{
    Algorithm to implement the plaquette operator $\nep^{i\gamma \calB}$ on the two plaquettes depicted in Fig.~\ref{fig:circuit_sketch}(c). The labeling of the qubit lines emphasizes that all the boundary qubits are shared with the neighboring plaquette pairs. The partially depicted green gates (dashed lines) are related to the simultaneous implementation of the same algorithm on a couple of neighboring plaquettes. They connect the displayed qubits with qubits belonging to the surrounding lattice sites, based on suitable translations of the two-plaquette unit.}
    \label{fig:twoplaquettes}
\end{figure}

We focus on a single set, which corresponds to our basic unit. We show in  Fig.~\ref{fig:twoplaquettes} the corresponding quantum circuit, that will be run \emph{in parallel} for all such sets.
Neighboring pairs of plaquettes share qubits at their boundary: this can be understood by ideally replicating the pair in Fig.~\ref{fig:circuit_sketch}(c), to build a lattice. For instance, the qubits 3 and 6 of our set also correspond to the qubits 2 and 5 of the set above the one in exam. Similarly, qubits 2 and 5 are homologous of qubits 3 and 6 for the set below, whereas qubit 7 matches qubit 1 of the plaquette pair lying on the right of the one depicted, and so on.

The algorithm defined in Fig. \ref{fig:twoplaquettes} performs the rotation of both plaquettes in 12 steps and it can be run in parallel for all plaquette pairs. 
This procedure is based on applying the phase gates $U_{p}(\gamma)$ on qubits 4 and 7 (thus on all the qubits of the lattice lying on the vertical links) to implement the plaquette rotation. The gates partially depicted in green are related to the simultaneous realization of the same algorithm for the neighboring plaquettes. All the qubits lying on the vertical links are required to be connected via CNOTs to the four neighboring qubits, whereas the horizontal qubits are just connected to two neighbors each.
As a result of the previous scheme, each of the P steps of the QAOA can be realized with a circuit of depth 13 on systems with open boundaries, or systems with closed boundaries and an even number of rows or columns. 
In this work, we focus most of the numerical investigation on a lattice with $3\times 3$ plaquettes and periodic boundary conditions. 
Its implementation on actual quantum hardware requires some additional care due to the boundary conditions and, in that case, each QAOA step can be realized with a circuit of depth 18 (see Appendix \ref{app:3x3}).

To evaluate the scaling of the preparation of a target ground state of the $\mathbb{Z}_2$ LGT, it is also important to consider the preparation of the initial states $\ket{\psi_0}$, which will subsequently be modified by the QAOA layers.
The electric ground state $\ket{\Omega_E}$ is a trivial product state and it can be prepared by applying Hadamard gates on all qubits, to rotate them into the eigenstate $\ket{+}$ of the $\PauliSigma^x$ operator.

The toric code (magnetic) ground state $\ket{\Omega_B}$ displays instead topological order and long-range entanglement \cite{Chen2010}.
To initialize this state, we follow the technique adopted in Ref.~\cite{Liu_arxiv2021}: assuming that each qubit is initially in the eigenstate $\ket{\uparrow}$ of $\PauliSigma^z$, in each plaquette we first apply three Hadamard gates on the control qubits and then three CNOT gates targeting the fourth  \footnote{Our implementation is equivalent to the procedure in Ref.~\cite{Liu_arxiv2021} up to a rotation of the basis. Hence, we have three control qubits and one target for each plaquette, instead of one control and three targets}.
This procedure is similar to the first 3-CNOT sequence of the circuit depicted in Fig.~\ref{fig:circuit_sketch}(a), with the addition of Hadamard gates on the qubits $\ket{p_1}$, $\ket{p_2}$, and $\ket{p_3}$.

These operations can be performed in parallel on plaquettes belonging to a single row and then repeated $L$ times to cover the whole lattice. 
Operations on different columns (or rows), however, cannot be parallelized because in each plaquette the CNOT gates must be applied before using one of the control qubits as the target for the neighboring column (row).
If we consider Fig.~\ref{fig:circuit_sketch}(c), the plaquette operations must start from the rightmost column, in such a way that the qubit $\ket{p_4}$ is used as a control before becoming the target of the plaquette on the left.

This is an important difference with respect to the application of the gate $\nep^{-i\gamma \Ham_B}$: despite the preparation of the ground state of $\Ham_B$ on a single plaquette requires a smaller number of gates, for large systems the initialization of $\ket{\Omega_B}$ requires a deeper circuit than the Hamiltonian gate, which, instead, can be run in parallel on all the even or odd columns (rows). 
This reflects the necessity of having a circuit with depth $O(L)$ to prepare a state with long range entanglement, such as $\ket{\Omega_B}$, which has been well studied in the literature~\cite{Bravyi_PRL2006,Chen2010, Bravyi_PRL2019}.  
In conclusion, when we compare the QAOA results with different choices of the initial state $\ket{\psi_0}$, we need to take into account the overhead required for preparing $\ket{\Omega_B}$.

The construction of the QAOA layers we presented so far was restricted to the case of an Abelian $\mathbb{Z}_2$ LGT. We stress, however, that the same procedure can be generalized to pure 2D LGTs with arbitrary discrete gauge groups. In particular, by suitably extending the Hilbert space associated with each link of the square lattice, it is possible to implement the time evolution steps of both the electric and magnetic Hamiltonian based on local unitary operators \cite{lamm}.

In this respect, the simplest generalization is provided by $\mathbb{Z}_n$ LGTs (see, for instance \cite{Horn1979,Zohar2013,Notarnicola2015,Zohar_PRA2017,Ercolessi2018,Emonts2020,Nyhegn2021,Robaina2021}). In this case, a gauge degree of freedom is encoded into an $n$-dimensional Hilbert space, as common in quantum clock models with $\mathbb{Z}_n$ symmetries \cite{fradkinkadanoff,Ortiz2012}. The electric field assumes indeed $n$ different values, which may be represented by suitable qudits (or by embedding each degree of freedom in a set of qubits). $\Ham_E$ remains a local Hamiltonian, whose time evolution can be perfomed in a parallel way over all links.

As in the case of the $\mathbb{Z}_2$ theory, also for $\mathbb{Z}_n$ symmetries there is a suitable unitary transformation mapping the eigenstates of the electric Hamiltonian into the eigenstates of the magnetic operators adopted to build the plaquette terms (the so-called connection operators). Such unitary transformations generalize the Hadamard gates we adopted and correspond to a quantum Fourier transform. Additionally, the plaquette term maintains the same 4-body interaction form through a suitable replacement of $\hat{\sigma}_z$ with quantum clock operators. The implementation of the plaquette operator thus requires generalizing the CNOT to controlled $\mathbb{Z}_n$ clock gates.
Again, the phase diagram of pure 2D $\mathbb{Z}_n$ LGT models presents a deconfined and topological phase at large $h$, whose topological order matches the $\mathbb{Z}_n$ generalization of the toric code \cite{Bullock_2007,Schulz_2012}, and a confined phase whose ground state becomes a trivial product for $h=0$.

By following the Kogut-Susskind Hamiltonian construction, a further generalization can be implemented to investigate ground states of discrete non-Abelian 2D LGTs (see, for example, Refs.~\cite{Burrello2015,Bender_NJP2018, lamm}). 
In this case, the gauge degrees of freedom can be represented either in an eigenbasis associated with the irreducible representations of the group, which is diagonal in the electric term of the Hamiltonian, or in an eigenbasis associated to the group elements, which is diagonal in the magnetic term of the Hamiltonian. The general structure of a quantum algorithm for implementing the QAOA steps in this case is analogous to the previous one and can be based on the construction in Ref.~\cite{lamm}. 
Given the non-Abelian nature of the group, however, the implementation of the magnetic time evolution requires a further technical generalization. 
In this case the irreducible representations are not one-dimensional and correspondingly the connection operators acquire a tensor form, therefore the gauge-invariant plaquette terms must be written in terms of their trace \cite{kogut1979}, requiring, in turn, to extend the rotation operators $U_{p}(\gamma)$ to more general single-link gates, which apply phases given by the traces of gauge group matrices.

\section{Numerical results}\label{sec:results}
In this section we analyze the QAOA performance on our LGT model, showing that the ground state can be prepared through shallow circuits with good fidelity both in the confined and in the topological phase.
Unless otherwise stated, 
the numerical analysis is performed on a lattice with $3\times 3 $ plaquettes (18 qubits) and implemented through the python package Qiskit~\cite{Qiskit}, using the circuit sketched in Sec.~\ref{ssec:circuit}.
Simulations of larger systems ($L=4,\ 5$), instead, exploit the mapping onto the 2D-TFIM to reduce the Hilbert space dimension and allow for the exact evaluation of the QAOA {\em Ansatz}. 

\subsection{Energy landscape}\label{ssec:energy_landscape}
The first feature we are interested in is the structure of the energy landscape, because it determines whether the classical optimization of the variational parameters can be performed efficiently or not~\cite{Cerezo_NatRev2021}.
Indeed, the presence of rugged energy landscapes is a common problem that severely affects the classical optimization loop of VQAs by making it prone to remaining stuck in local minima, some of which might be far from the ground state energy.
\begin{figure}
    \centering
    \includegraphics[width=8.5cm]{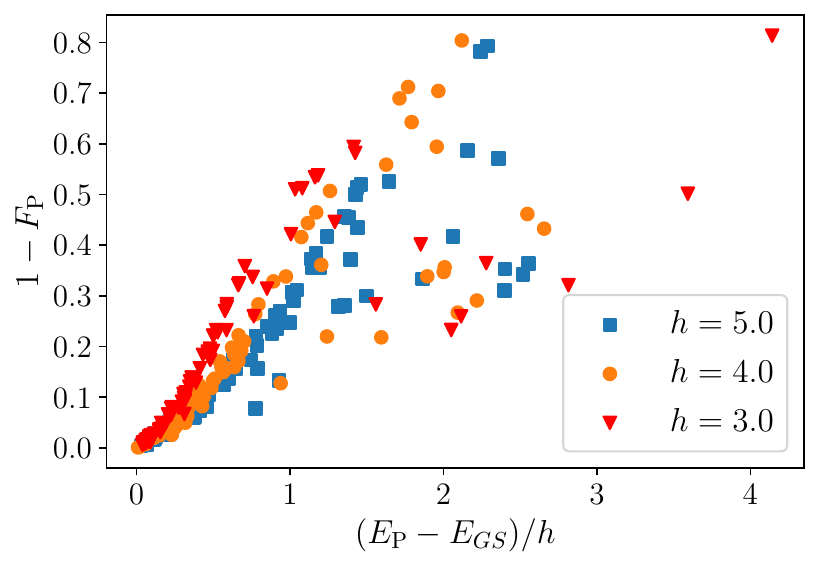}
    \caption{Infidelity vs residual energy rescaled over the magnetic coupling $h$, for $h=5,\ 4, \ 3$. Data refer to 100 local optimizations on a system with linear dimension $L=3$, initial random guess of the QAOA parameters, and circuit depth $\Ptrot=5$. 
    The initial state is $|\psi_0\rangle=|\Omega_E\rangle$.}
    \label{fig:randominit}
\end{figure}
To characterize the energy landscape and to quantify the quality of the optimized variational {\em Ansatz} $\ket{\psi_\Ptrot(\bgamma^*,\bbeta^*)}$, we use either the residual energy or the fidelity.
Given a target Hamiltonian $\Ham_\target(h)$, we denote with $\ket{\psitarget }$ its ground state and with  $E_{GS}$ the corresponding energy, both obtained with exact diagonalization.
The residual energy is simply the difference between the minimized variational energy $E(\bgamma^*, \bbeta^*)$, defined in Eq.~\eqref{eq:QAOA_en}, and $E_{GS}$, while the fidelity follows the usual definition 
\begin{equation}
    F_\Ptrot(\bgamma^*,\bbeta^*) =
|\braket{\psi_\Ptrot(\bgamma^*,\bbeta^*)}{\psitarget}|^2 \ . 
\end{equation}
In this work, we use the fidelity as a precise estimate of the accuracy of the approximation of the target state, even though, in an actual experiment, it is hardly accessible because it requires an exponential number of measurements.
The energy, instead, is easily estimated: the magnetic contribution is diagonal in the computational basis, while the electric contribution is obtained by applying a basis rotation on each qubit, i.e. a set of Hadamard gates, before the measurement.

In the $\Ztwo$ LGT model, the energy landscape emerging from the QAOA {\em Ansatz} is characterized by many local minima covering a wide energy interval, making random-start local optimization impractical. 
This effect is particularly severe if the target state is in a phase different from the initial one $|\psi_0\rangle$. This is illustrated in Fig.~\ref{fig:randominit}, where we show the residual energy and the infidelity $1-F_\Ptrot(\bgamma^*,\bbeta^*)$ 
for 100 different random-start local optimizations with $\Ptrot=5$ QAOA layers and three different values of the magnetic coupling, around or above the topological phase transition. 
The initial state is the product state $\ket{\Omega_E}$ in the extreme confined limit $(h=0)$ and the local minimizations are performed with the Broyden–Fletcher–Goldfarb–Shanno (BFGS) algorithm~\cite{Nocedal_book2006}.
Although there is a clear concentration of data in the corner corresponding to successful optimizations, where both the infidelity and the residual energy tend to zero, there are many local minima far from the ground state, suggesting that 
more refined optimization techniques are needed for this model.

However, the clear correlation between the energy and the fidelity is reassuring since the absence of low-energy minima with small projection on the ground state guarantees that any scheme that allows for a reliable minimization of the energy will also lead to a good approximation of the target state. This correlation is intuitively justified by the existence of a gap in the topological phase. Indeed, the construction in Eq. \eqref{eq:QAOA_state} cannot mix different topological sectors of the model and, for each topological sector, there is only one ground state. Away from the critical point the ground state is protected by a finite gap, such that the correlation between infidelity and residual energy must hold below this energy scale. Close to the critical point, other orthogonal low-energy states may appear and spoil the correlation. This, however, seems not to be the case. Its resilience is not surprising for small system sizes in which the gap does not close even at $h_c$. However, we observe that the correlation between infidelity and residual energy holds also when we increase the linear dimension $L$ of the lattice, and it actually appear to be even sharper, as shown in App.~\ref{app:energy}.
The variational energy is therefore a reliable figure of merit for the optimization, it can be efficiently measured in experiments, and the procedure is still effective
when the system size is scaled up. All these represent positive factors for the feasibility of the implementation of QAOA on the $\Ztwo$ LGT model in near term quantum devices.  

\subsection{Heuristic local optimization: two-step QAOA}\label{sec:results_ts}
Because of the large number of suboptimal minima present in the energy landscape, it is important to adopt an efficient strategy in order to reliably find a good approximation of the true ground state of the target Hamiltonian.
This is, indeed, a crucial task for QAOA and VQAs in general, where the classical optimization outer-loop is often the main computational bottleneck and several strategies have been proposed that go beyond a local search from a random start.
These strategies range from problem-specific methods to general iterative techniques, based on observed patterns in the optimal schedules~\cite{Zhou_PRX2020,Wauters_PRR2020, Streif_2020,Wurtz_PRA2021}. 

We adopted here a simple two-step minimization protocol inspired by 
a digitized Quantum Annealing \cite{Mbeng_PRB2019,Mbeng_arx19} turning-on of one of the two terms of the Hamiltonian.
The idea behind the two-step optimization is to leverage on the formal analogy between QAOA and digitized Quantum Annealing (QA)~\cite{Mbeng_arx19}: for depth-$\Ptrot$ QAOA, we first optimize the total run time of a digitized linear QA \cite{Mbeng_PRB2019} of the same depth, and then fine-tune the variational parameters around this schedule.
This approach can be used effectively when the system is initialized either in the electric state $\ket{\Omega_E}$ or in the magnetic one $\ket{\Omega_B}$. 
For generic applications, the confined electric ground state $\ket{\Omega_E}$ --- a uniform superposition of all possible states in the computational (magnetic) basis --- is a standard choice for initializing the variational circuit, because it is easy to prepare.
However, this choice is non-optimal when we target states with long-range entanglement in the deconfined/topological phase, since a circuit of local unitary gates with bounded depth cannot create states with topological order beyond a certain system size~\cite{Chen2010}.

When the initial state is set to be $\ket{\Omega_E}$, an adiabatic turning-on in a time $\tau$ of the magnetic coupling through $\Ham(t)=\Ham_E + (t/\tau) h\Ham_B$ suggests --- after digitization by Trotter decomposition of $\nep^{-i\Delta t \Ham(t_m)}\approx \nep^{-i\Delta t\Ham_E}\nep^{-i \frac{m\Delta t}{\Ptrot} h\Ham_B}$, with $t_m/\tau=m/\Ptrot$ --- setting $\gamma_m^0 = \frac{m\, \Delta t}{\Ptrot}h$ and $\beta_m^0 = \Delta t$ in the state in Eq.~\eqref{eq:QAOA_state}.
In our first QAOA step, these linear-schedule parameters are optimized by searching for the optimal digitized QA~\cite{Mbeng_PRB2019} $\Delta t^*$ --- a simple one-dimensional minimization --- which leads to setting:
\begin{equation} \label{eq:dQA_el}
    \gamma_m^{\dQA} = \frac{m\, \Delta t^*}{\Ptrot} h \;, \hspace{5mm}  \beta_m^{\dQA} = \Delta t^* \;. 
\end{equation}
The second step in our QAOA procedure is to perform 10 local BFGS optimizations in the $2\Ptrot$-dimensional parameters space, starting from $(\bgamma^{\dQA},\bbeta^{\dQA})+\bepsilon$, 
where $\bepsilon$ is a small $2\Ptrot$-dimensional vector with random numbers uniformly distributed in the interval $[-0.025,0.025)$, 
keeping the best outcome out of these local optimizations.
Schematically:
\begin{equation} \label{eq:scheme_QAOA}
(\bgamma^{\dQA},\bbeta^{\dQA}) + \bepsilon
\hspace{3mm}\stackrel{\scriptscriptstyle \mathrm{best \, BFGS}}{\longrightarrow} \hspace{3mm} (\bgamma^\star,\bbeta^\star) \;.
\end{equation}

The toric code ground states $\ket{\Omega_B}$, corresponding to the extreme deconfined limit $h\to \infty$, provide a better initial state $\ket{\psi_0}$ when targeting ground states in the topological phase: they can be exactly prepared with local circuits whose depth scales with the width of the system ~\cite{Satzinger_Science2021,Liu_arxiv2021}.
Proceeding once again with an adiabatic turning-on, now of the electric part of the Hamiltonian, through $\Ham(t)=h(\Ham_B+(t/h\tau)\Ham_E)$, suggests --- after digitization by Trotter decomposition of $\nep^{-i(\Delta t/h) \Ham(t_m)}\approx \nep^{-i \Delta t \Ham_B} \nep^{-i\frac{m\Delta t}{h\Ptrot} \Ham_E}$, with $t_m/\tau=m/\Ptrot$ --- setting $\gamma_m^0 = \Delta t$ and $\beta_m^0 = \frac{m\, \Delta t}{h\Ptrot}$ in the state in Eq.~\eqref{eq:QAOA_state}.
Once again, these can be optimized by searching for the optimal digitized QA $\Delta t^*$, which leads to:
%
\begin{equation} \label{eq:dQA_magn}
    \gamma_m^{\dQA} = \Delta t^* \;, \hspace{5mm}
    \beta_m^{\dQA} = \frac{m\, \Delta t^*}{h\, \Ptrot} \;.
\end{equation}
The second step in our QAOA procedure is identical to the previous case, as schematically indicated in Eq.~\eqref{eq:scheme_QAOA}.

Two noteworthy features of the QAOA minima obtained by applying our two-step QAOA procedure are the {\em smoothness} of the schedules $(\bgamma^\star,\bbeta^\star)$, illustrated in Appendix~\ref{app:smooth}, and the closely related {\em transferability} of such smooth schedules from a smaller to a larger $L'>L$ sample, discussed in Sec.~\ref{sec:transferability}.
We benchmarked our heuristic two-step QAOA approach against a computationally expensive global optimization, finding comparable quality results in terms of ground-state fidelity, both in the confined and in the deconfined phase: we illustrate this in Appendix~\ref{app:globalvslocal}.

In the following, we will compare the performance of our two-step QAOA for systems prepared either in the electric ground state $\ket{\Omega_E}$, or in the toric code ground state $\ket{\Omega_B}=\ket{++}$. 
For a fair comparison, a remark is in order: while $\ket{\Omega_E}$ is trivially prepared with one layer of single-qubit Hadamard gates, for the preparation of $\ket{\Omega_B}$ one should include an overhead circuit with $O(3L^2)$ gates, organized in $L$ layers applied sequentially.
As explained in Sec.~\ref{sec:alg}, although no optimization is necessary for this preliminary step, it is still required to apply 3 CNOT gates for each plaquette.

\begin{figure}
    \centering
    \includegraphics[width=8.5cm]{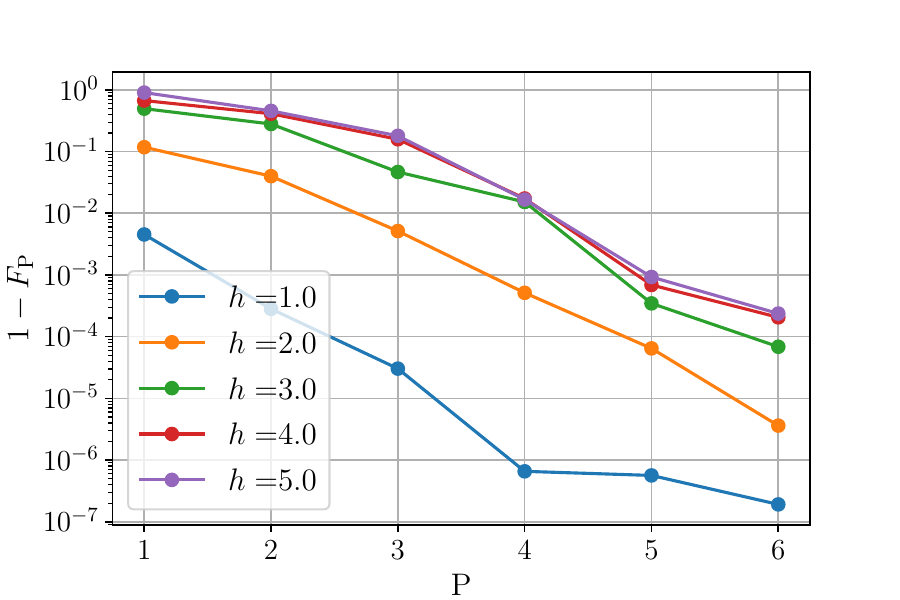}
    \caption{Infidelity vs the number of QAOA layers $\Ptrot$ for $L=3$ and  several values of the magnetic coupling $h$.
    The initial state is the electric ground state $\ket{\Omega_E}$, thus the convergence towards the exact GS is faster for smaller values of $h$.
    We show the best result out of ten local BFGS minimizations following the heuristic two-step optimization method.}
    \label{fig:QAOA_opt}
\end{figure}

The QAOA results obtained from the initial product state $\ket{\Omega_E}$ are reported in Fig.~\ref{fig:QAOA_opt}, where we show the infidelity $1-F_\Ptrot(\bgamma,\bbeta)$ as a function of the circuit depth $\Ptrot$ for several values of the magnetic coupling, both in the confined phase ($h \lesssim 3)$ and in the deconfined one $(h\gtrsim 3)$.
As expected, the variational {\em Ansatz} converges faster to states in the same phase (e.g. $h=1,\ 2$) but QAOA can reach very good fidelity $1-F_\Ptrot < 10^{-3}$, when $\Ptrot \ge 5$, for all the couplings we considered.

\begin{figure}
    \centering
    \includegraphics[width=8.5cm]{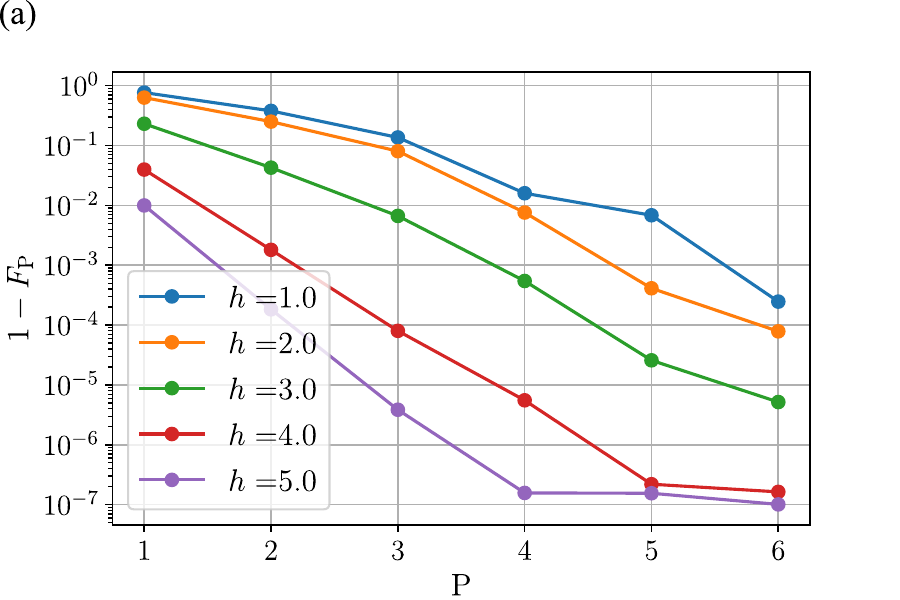}
    \includegraphics[width=8.5cm]{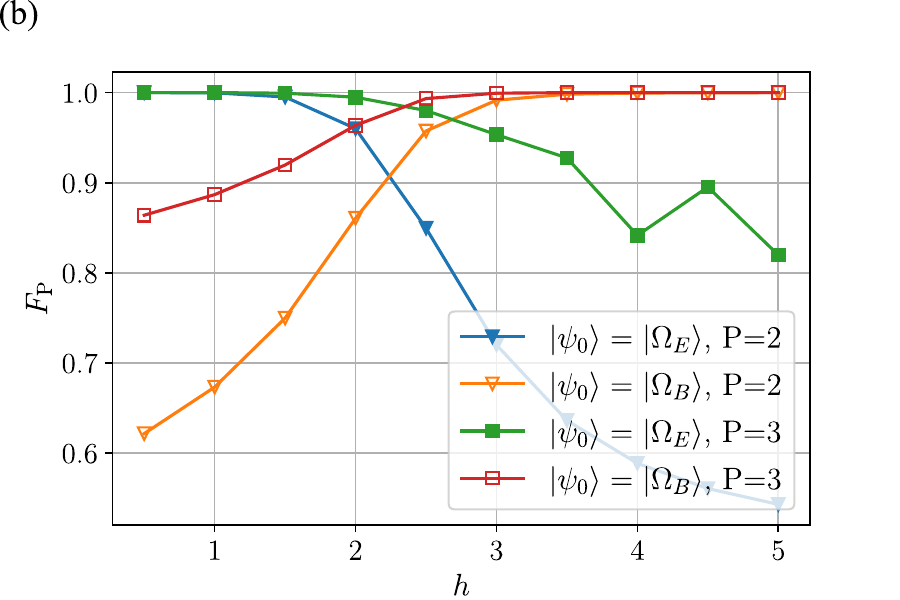}
    \caption{(a): Infidelity vs the number of QAOA layers $\Ptrot$ for $L=3$ and  several values of the magnetic coupling $h$.
    Data correspond to the best out of 10 results obtained in the two-step optimization, performed on a state initially prepared in the toric code ground state $\ket{\Omega_B}$, hence the convergence is now faster for larger couplings $h$. 
    (b): Comparison between two-step QAOA performance by starting from the electric and the magnetic ground states: we plot the fidelity vs magnetic coupling $h$ at fixed values of $\Ptrot$.
    Here $\Ptrot$ only takes into account the number of QAOA layers with parameterized gates, while it does not include the 
    computational overhead for the preparation of
    $\ket{\Omega_B}$, compared to preparing $\ket{\Omega_E}$.  
    }
    \label{fig:QAOA_opt_reverse}
\end{figure}

With the ``reversed'' protocol, starting from the toric code ground state $\ket{\Omega_B}=\ket{++}$, we obtain an overall behavior similar to what observed for $\ket{\psi_0}=\ket{\Omega_E}$, see  Fig.~\ref{fig:QAOA_opt_reverse}(a), with the important difference that now the optimization converges faster when targeting the deconfined phase.
Indeed, only $\Ptrot=3$ QAOA layers are now needed to reach $1-F_\Ptrot < 10^{-3}$ when $h>h_c$, see data for $h=4$ or $h=5$, 
while confined states require more QAOA layers to reach comparable accuracy.

For both choices of initial state, we observe that the infidelity decreases exponentially with the circuit depth; the only exceptions for $\Ptrot=5,6$ can be ascribed to the algorithm remaining stuck in a (high-quality) local minimum, when the target state is very close to the initial one (see Appendix~\ref{app:smooth}). 
However, if we focus on the minimal resources to approximate the target state within a certain fidelity threshold, we can further reduce the number of parameters required.
Figure~\ref{fig:QAOA_opt_reverse}(b) shows a comparison of QAOA performance with the two possible choices of the initial state, for $\Ptrot = 2$ and $\Ptrot= 3$, by looking at the best fidelity reached by the two-step optimization as a function of the coupling $h$.
Remarkably, such shallow variational circuits are enough to prepare with high fidelity the ground states in the confined and deconfined phases, provided the initial state is selected in the same phase as the target ground state. 
Unsurprisingly, the region that requires a larger number of parameters, i.e. a deeper variational circuit, corresponds to the crossover between the two regimes, where $2 \lesssim h \lesssim 3$.

We finally observe that the choice of the initial state based on the target value of $h$ plays a role analogous to the choice of the electric or magnetic representation of the LGT Hamiltonians applied in the quantum simulation protocols presented in Refs. \cite{Haase2021,Paulson_PRX2021}.

\subsection{Schedule transferability} \label{sec:transferability}
A promising route to reduce the computational cost of the outer-loop classical optimization in VQAs is the \emph{transferability} of the optimal parameters from small to large instances of the same model.
Indeed, as empirically observed or proven in specific applications of VQAs, if you consider two instances of the same model and a fixed variational circuit depth $\Ptrot$, the optimal parameters obtained for the small system of size $L$ may serve as a very good warm-start (or educated guess) for a local optimization for the $L'$-size model ($L'>L$)~\cite{brandao2018, farhi2020quantum, Grover_concentration}.

Classical numerical simulations soon become unfeasible even for modest sizes, often hindering a more systematic analysis on this issue: for our LGT model, which requires $2L^2$ quantum spins, even sizes as small as $L=4,5$ can be extremely challenging to simulate exactly.
To partially overcome the size limitation, we exploit the mapping onto the 2D-TFIM, explained in Sec.~\ref{sec:model}, which involves only $L^2$ spins on a square lattice, by taking advantage of the restrictions imposed by the gauge constraints. 
This allows us to simulate exactly the variational optimization for $L=4,5$.
To study the schedule transferability, we first perform the two-step QAOA on the system with $L=3$, as described in Sec.~\ref{sec:results_ts}.
The optimal angles $(\bgamma^\star, \bbeta^\star)$ found for $L=3$  are then used as warm-start points for a local optimization on larger sizes.
In particular, we keep the best run out of 10 BFGS optimizations on the larger systems, each of them starting in the neighborhood of $(\bgamma^\star, \bbeta^\star)_{L=3}$, similarly to the strategy used in the second part of the two-step QAOA protocol.
This procedure is repeated for different values of the coupling.

\begin{figure}
    \centering
    \includegraphics[width=8.5cm]{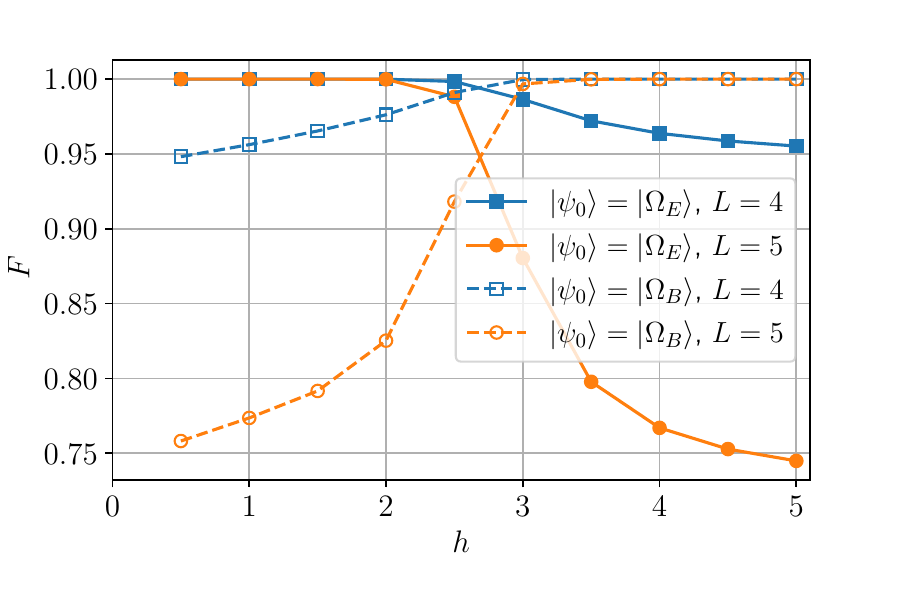}
    \caption{Fidelity vs the magnetic coupling $h$. The data are obtained by using the two-step optimal schedules for $L=3$, as initial guess for 10 local BFGS optimizations performed on the larger systems $L=4,5$ (we fixed $\Ptrot=6$). Here, we report the best result out of the 10 runs. 
    The full and empty symbols respectively correspond to 
    $\ket{\psi_0}=\ket{\Omega_E}$ and $\ket{\psi_0}=\ket{\Omega_B}$.}
    \label{fig:Scalability}
\end{figure}

The results obtained are reported in Fig.~\ref{fig:Scalability}, where we compare the fidelity $F_\Ptrot(\bgamma^*,\bbeta^*)$ vs $h$, for circuit depth $\Ptrot=6$ --- which allows us to prepare the ground state for arbitrary $h$ with an error $1-F_\Ptrot(\bgamma^*,\bbeta^*)<10^{-3}$ for $L=3$ --- and both possible initial states: $\ket{\psi_0}=\ket{\Omega_E}$ (full symbols) and $\ket{\psi_0}=\ket{\Omega_B}$ (empty symbols).
The transferability of the parameters is almost perfect when the initial and target states are in the same topological phase, leading to very high fidelities both in the small and large magnetic coupling regimes.
Even when we target a ground state in a different phase than the initial one --- for instance, $\ket{\psi_0}=\ket{\Omega_E}$ and $h>h_c\sim 3.0$ ---  the final fidelity is still large, allowing us to characterize the topological properties of the final state.
The poorer fidelity observed for $L=5$, when targeting a state in the opposite phase than $\ket{\psi_0}$, is due only to the reduced ratio between the circuit depth $\Ptrot$ and the size $L$ (particularly relevant when crossing a topological phase transition \cite{Chen2010}) and not to the size of the system on which the original schedule has been optimized ($L=3)$.
Indeed, transferring the schedule from $L=4$ to $L=5$, or from $L=3$ to $L=5$, leads to the same final performance.

Interestingly, the warm-start initialization provided by the $L=3$ optimal parameters leads to a successful \emph{local} minimum search for $L=4,5$,
with an accuracy close to what can be achieved with a full global minimization, discussed in Appendix~\ref{app:globalvslocal}. 
Moreover, the number of iterations needed for the local optimization is rather small ($N_{\rm iter} \lesssim 50$), confirming the benefit of the transferability of optimal solutions: once the $L=3$ two-step solution is provided, only a small overhead in computation resources is required to fine-tune the parameters for larger sizes. 
Hence, transferability provides a speed advantage over starting from scratch a two-step optimization: even though the fidelity reached is comparable, the latter requires more runs of the quantum circuit, making it less efficient when the optimal schedule for a smaller system is already known.

This transferability evidence may be linked to the observation of the smooth schedules we found with the two-step optimization, as shown in Appendix~\ref{app:smooth}.
It is also important to remark that the schedule transferability is not a general property of any minimum in the energy landscape, but it is associated with the smooth solution found with the two-step protocol.
For instance, a global optimization yields slightly better results on the $L=3$ system, but it often represents a poor choice as an educated guess to initialize a local minimum search on larger sizes, as discussed in Appendix~\ref{app:globalvslocal}: this phenomenon is similar to \emph{overfitting} in machine learning~\cite{Goodfellow_ML:book}.
In this respect, the two-step scheme appears to outperform an extensive global search.

\subsection{Ground state characterization}
In the following, we turn our attention to the properties of the approximate ground states we prepare with QAOA.
Despite the finite size limitations of our simulations, the states obtained through QAOA display most of the main features associated with the appearance of topological order and the crossover from a confined to a deconfined phase as $h$ increases.
The main observables to distinguish these two regimes are the Wilson loops, as defined in Eq.~\eqref{eq:Wilson}. 
We consider in particular Wilson loops $\mathcal{W}_{l_x,l_y}$ defined over rectangles of size $l_x \times l_y$.

\begin{figure}
    \centering
    \includegraphics[width=8cm]{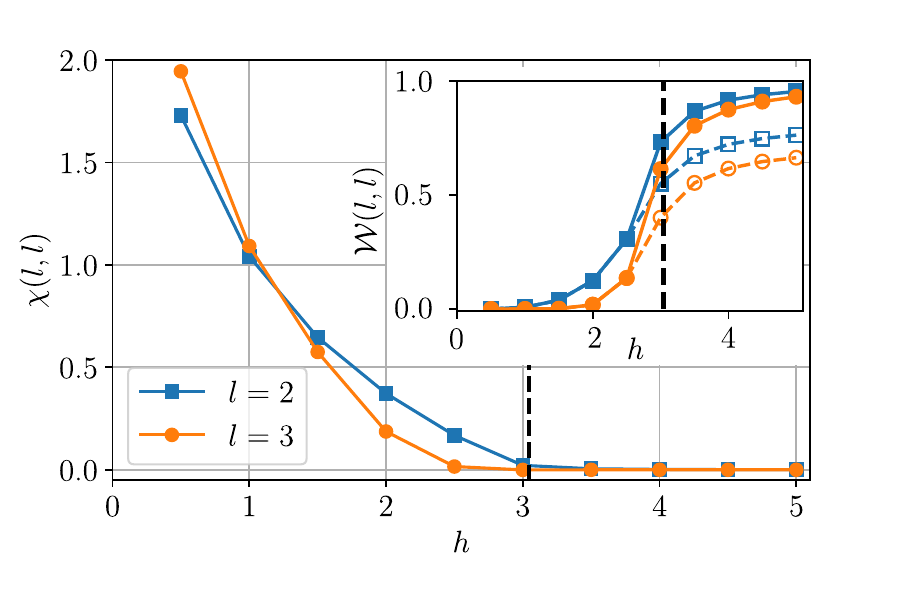}
    \caption{Creutz ratio $\chi(l,l)$, defined in Eq.~\eqref{eq:Cruetz_ratio}, for two different loops in a system with $L=5$.
    Inset: expectation value of Wilson operators $\mathcal{W}_{l,l}$, corresponding to the data in the main plot, vs the coupling strength $h$. The vertical dashed line indicates the critical value of the coupling $h_c$.
    All data refer to  the best energy out of 20 BFGS local optimizations, with $\Ptrot=6$ and $\ket{\psi_0}=\ket{\Omega_E},\ket{\Omega_B}$, performed on a system of linear size $L=5$ and initialized with the optimal parameters found for $L=3$.
    Empty symbols in the inset are data obtained with $\ket{\psi_0}=\ket{\Omega_E}$.
}
    \label{fig:wilson_g}
\end{figure}

As explained in Sec.~\ref{sec:model}, it is known that the deconfined phase is characterized by an exponential decay of $\langle \mathcal{W}_{l_x,l_y} \rangle$ with the perimeter $P$ of the loop, whereas the confined phase displays a decay dictated by the area $A$ of the loop \cite{Greensite2003}. In particular, the magnetic ground states $\ket{\Omega_B}$ are such that $\langle \mathcal{W}_{l_x,l_y} \rangle=1$, while in the electric ground state $\ket{\Omega_E}$ Wilson loops always present vanishing expectation values. Recalling Eq.~\eqref{eq:wilson_gen}, the overall behavior of a Wilson loop can be approximated by  $\langle \mathcal{W} \rangle \propto \nep^{-\chi A - \delta P}$.
Indeed, if $\chi > 0$, the exponential decay with the area dominates for large loops, while if instead $\chi=0$, the decay is dictated by the perimeter law only.
To extract the information about the $\chi$ coefficient we estimate the so-called Creutz ratio \cite{Creutz_PRL1980}:
\begin{equation}\label{eq:Cruetz_ratio}
    \chi(l,l) = -\log \frac{\langle \Wilson{l, l} \rangle\langle \Wilson{l-1, l-1} \rangle}{\langle \Wilson{l, l-1} \rangle\langle \Wilson{l-1, l} \rangle}\,.
\end{equation}
This ratio is indeed built to cancel the perimeter contribution to the decay of the observables and approximate the coefficient $\chi$, which is recovered for large $l$. 

Fig.~\ref{fig:wilson_g} displays the Creutz ratio in a system with $L=5$ and periodic boundary conditions for states obtained with $P=6$ QAOA steps applied either to the state $\ket{\Omega_E}$ (for $h < 3$) or to $\ket{\Omega_B}$ (for $h \ge 3$). 
The optimization on the $L=5$ systems was initialized with the best result obtained with the two-step protocol for $L=3$, on top of which we performed 20 local minimum searches, out of which we consider the most successful outcome. 

Analogously to other LGT studies on small lattices \cite{Zohar2015fermionic,Bender2020,Zohar_PRR2021}, the finite size effects in our computation are  strong. When considering a Wilson loop of width 3, its opposite sides lay at distance 2. This implies that what we observe in Fig. \ref{fig:wilson_g} may provide a quantitative estimate of the behavior in thermodynamic systems only if the correlation length is sufficiently smaller than this distance, thus only sufficiently far from the phase transition. Despite this limitation, the Creutz ratio $\chi(3,3)$ presents a behavior that clearly distinguishes the confined phase ($\chi>0$) and the deconfined phase ($\chi \to 0$) appearing for $h\gtrsim 3$, although a quantitative identification of $h_c$ is beyond the possibilities of these small systems and loops. 

The inset of fig.~\ref{fig:wilson_g} reports the expectation value of the Wilson loop operators corresponding to the Creutz ratios shown in the main plot.
It clearly shows a crossover between the trivial, confined state with $\langle \Wilson{\Gamma}\rangle \to 0$ and the topologically ordered, deconfined limit $\langle \Wilson{\Gamma}\rangle \to 1$.
With the chosen scheme, i.e. starting from $\ket{\Omega_E}$ or $\ket{\Omega_B}$ depending on the target state, they perfectly match the results from exact diagonalization (not shown) as expected from the high fidelity reached, see Fig.~\ref{fig:Scalability}.  We emphasize that the possibility of obtaining a reliable estimate of the expectation value of the Wilson loops yields further important implications: 
 Ref. \cite{Paulson_PRX2021} shows indeed that, in a $U(1)$ LGT, even the expectation value of the single plaquette operator can be used to extract the running coupling of the model, which is a fundamental quantity related to its renormalization.

If we chose to start always from the electric ground state, the deviation from exact results would become larger in the deconfined phase, as also expected from the fidelity drop observed in Fig.~\ref{fig:Scalability}.
However, the results obtained in this non-optimal case still provide an acceptable scaling of the Wilson loop for the deconfined regime (empty symbols in Fig.~\ref{fig:wilson_g}): even without perfect reconstruction of the target state, it is still possible to identify the deconfined phase.
This is, indeed, useful for experimental investigation, where realistic setups are limited to shallow circuits and noise would decrease the quality of the approximated ground state.

Another observable that marks the onset of topological order is the topological entropy \cite{Kitaev2006,Levin2006}.
Given a connected subsystem $A\cup B \cup C$ of the whole lattice, we define its topological entropy as in the equation below
\begin{equation}\label{eq:Stopo}
    S_{topo} = S_A + S_B + S_C - S_{AB} - S_{BC} - S_{AC} + S_{ABC} \ .
\end{equation}
Here $S_X$ is the von Neumann entanglement entropy of a generic subsystem $X$, obtained by tracing out all degrees of freedom in the complement of $X$ with respect to the whole system, and $\lbrace A,B,C \rbrace$ is a tripartition of the region of which we compute the topological entropy.
In the toric code state $\ket{\Omega_B}$, the topological entropy of any subsystem is $S_{topo}=-\ln{2}$ and the total entanglement entropy is 
\begin{equation}
    S_{ABC} = N_v \ln{2} + S_{topo} = (N_v-1) \log{2} \ ,
\end{equation}
where $N_v$ is the number of vertex operators $\calA_v$ cut by the edge of the bipartition $X$ \cite{Hamma2005,Satzinger_Science2021}. 
In a product state, such as $\ket{\Omega_E}$, we expect both quantities to be zero, while for generic values of $h$ the entropy should interpolate between the two limits.
\begin{figure}
    \centering
    \includegraphics[width=4.5cm]{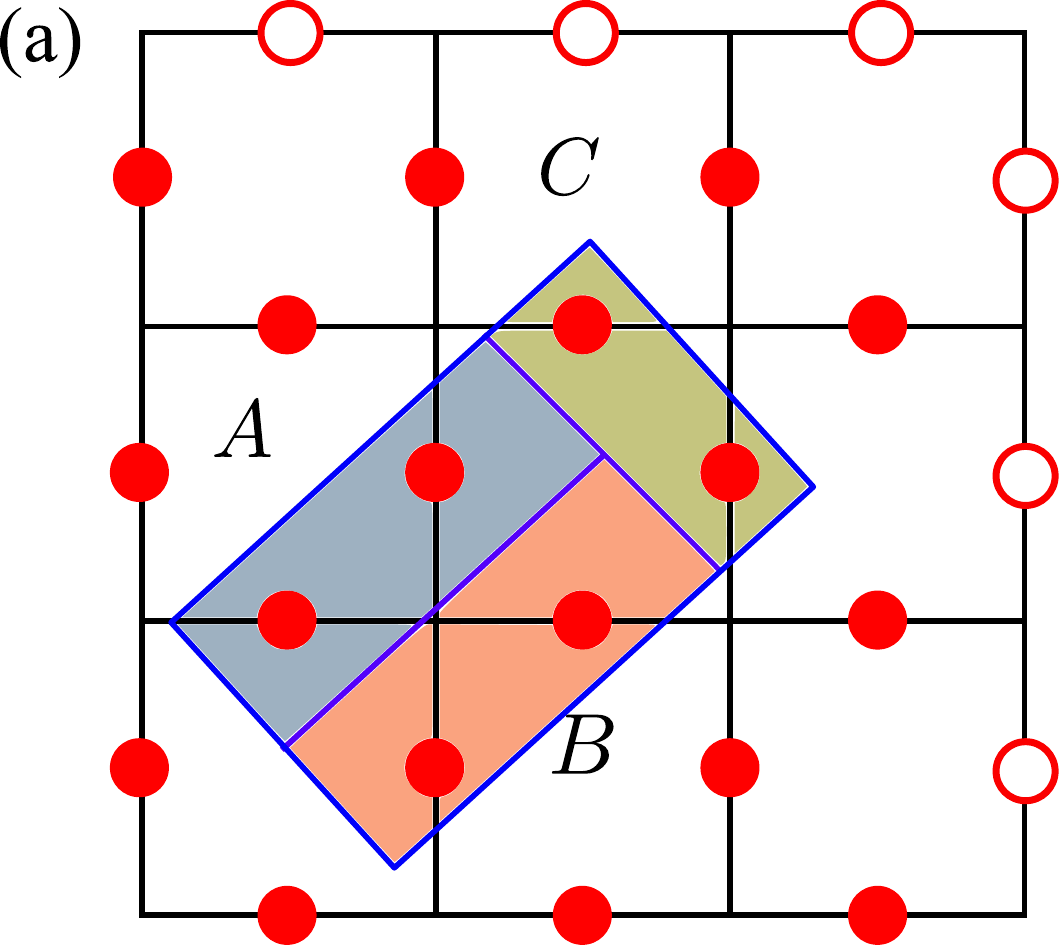}
    \includegraphics[width=8.5cm]{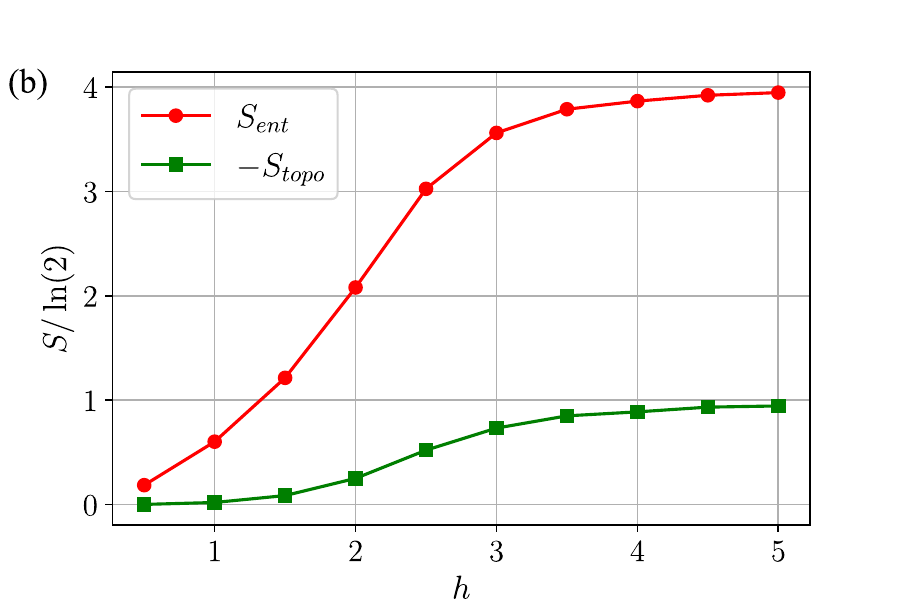}
    \caption{(a): Graphical representation of the subsystem used to compute the topological and entanglement entropy, with the tripartition $A,\ B, \ C$ highlighted. Notice that a total of $N_v=5$ vertex are cut by the outer edge of $A \cup B \cup C$. Empty dots indicate the presence of PBC in the lattice, thus identifying the upper edge with the lower one and the right edge with the left one.
    (b): Entanglement and topological entropy as a function of the coupling $h$. Notice that we plot $-S_{topo}$ to make it positive.}
    \label{fig:entropy}
\end{figure}
To compute the entropy, we choose a subsystem $X$ with 6 qubits, as depicted in Fig.~\ref{fig:entropy}(a), and we divide it into three further regions $A$, $B$, and $C$ with two qubits each.
We compute the entanglement entropy of all the subsets used in Eq.~\eqref{eq:Stopo} by tracing out explicitly their complements and obtain the data plotted in Fig.~\ref{fig:entropy}(b).
Despite the small dimension of the lattice and its subsystem, our results agree perfectly with the theoretical prediction: in the deconfined phase, the total entanglement entropy is $S_{\rm ent}=4\ln{2}$, since the partition $ABC$ cuts five vertices and the topological entropy approaches $S_{\rm topo}=-\ln{2}$.

\begin{figure}
    \centering
    \includegraphics[width=8.5cm]{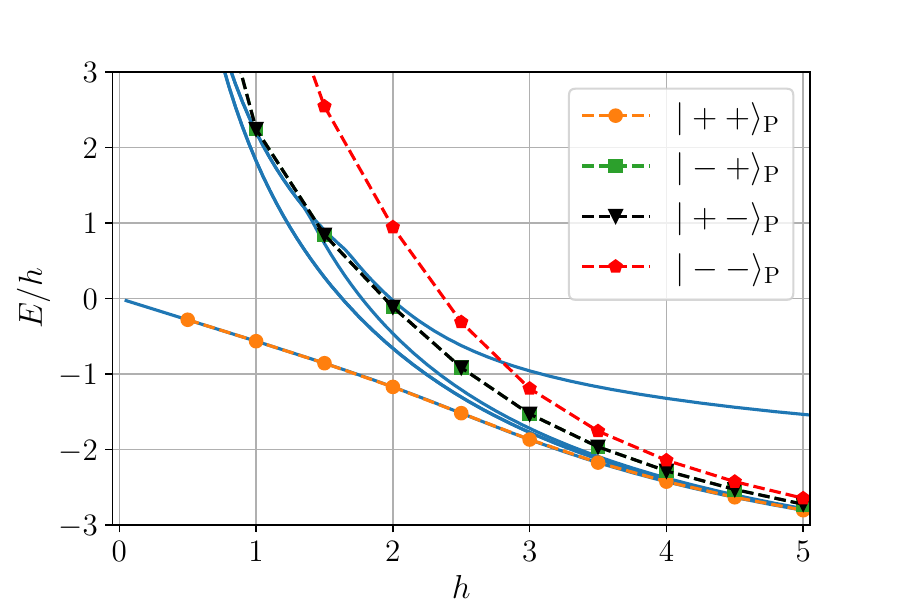}
    \caption{Four lowest energy states in different symmetry sectors of the non-contractible 't Hooft loop operators vs coupling strength $h$.
    All data refer to the best solution found for $L=3$, $\Ptrot=6$, and $\ket{\psi_0}=\ket{\Omega_E}$. States different from $\ket{++}_\Ptrot$ are obtained by acting with non-contractible Wilson loops after the unitary evolution, while the subscript $\Ptrot$ indicates that they are obtained by QAOA and thus are not exact eigenstates.
    The solid lines correspond to the 5 lowest eigenvalues obtained with exact diagonalization: The 5th eigenvalue, corresponding to the first excited state, is shown to highlight the topological gap in the deconfined phase. 
    Only four lines are visible because $\ket{+-}$ and $\ket{-+}$ are exactly degenerate, as their approximations obtained with QAOA.
    }
    \label{fig:bestresults}
\end{figure}

Finally, we would like to show that it is possible to manipulate the state constructed with QAOA to change its symmetry sector when the system has PBC.
Let $\ket{++}_\Ptrot$ denote the approximate ground state constructed with a QAOA circuit of $\Ptrot$ layers. We then construct approximate candidate ground states in the other topologial sectors by applying non-contractible Wilson loops $\Wilson{v/h}$, i.e., $\ket{+-}_\Ptrot=\Wilson{h}\ket{++}_{\Ptrot}$, $\ket{-+}_\Ptrot=\Wilson{v}\ket{++}_{\Ptrot}$, $\ket{--}_\Ptrot=\Wilson{h}\Wilson{v}\ket{++}_{\Ptrot}$. 
(The subscript label $\Ptrot$ is here used to distinguish the states obtained via QAOA from the exact eigenstates of the Hamiltonian.)
Non-contractible Wilson loops $\Wilson{v/h}$ are immediately implemented via $L$ single-qubit gates $\PauliSigma^z$ acting on a vertical or horizontal line. 
By doing so, however, we introduce an extra error on top of the finite accuracy of the QAOA state:
Indeed, choosing a specific vertical or horizontal Wilson loop to change the symmetry sector of the system breaks the translational invariance of the constructed state, producing a small excitation. This effect is visible in  Fig.~\ref{fig:bestresults}, where we show the energies of the state approximated with QAOA, denoted by $\ket{++}_\Ptrot$, and of the other three states obtained by applying $\Wilson{h}$ and $\Wilson{v}$ on $\ket{++}_\Ptrot$.
For comparison, we plot also the low eigenvalues obtained by exact diagonalization (drawn with solid blue lines).
For large $h$, the four lowest energy levels should be almost degenerate and, indeed, the exact diagonalization results are almost indistinguishable for $h\ge 4$.
In the same region, the excess energy of the approximate states $\ket{\tau_h,\tau_v}$ is instead clearly visible, although well below the topological gap with  the first proper excited state.

An alternative procedure to explore the different topological sectors in the deconfined regime, is to apply {\em first} the relevant Wilson loop on $\ket{\Omega_B}$ and then the QAOA unitaries.
In such a way, the initial state $\Wilson{h/v}|\Omega_B\rangle$ is {\em exactly} degenerate with $\ket{\Omega_B}$.
We find that the optimal schedule $(\bgamma^*,\bbeta^*)$ used to prepare the state $|++\rangle_{\Ptrot}$ minimizes also the expectation value of the energy in the other topological sectors, so no further optimization is required.
However, the picture presented in Fig.~\ref{fig:bestresults} remains valid and small excitations are created in the other topological sectors.
In other words, by inverting the order of application of the operators $\Wilson{h/v}$ and $\widehat{U}(\bgamma^*,\bbeta^*)$, we observe nearly irrelevant changes on the expectation value of the energy; $\widehat{U}(\bgamma^*,\bbeta^*)$ is the QAOA evolution operator with optimal parameters for the state $\ket{++}_\Ptrot$.

The expectation value of the 't Hooft loops $\tau_h$ and $\tau_v$, which distinguish the different topological sectors, is perfectly reconstructed by the algorithm.
This last feature is, however, independent from the specific values of $h$ and $\Ptrot$, since the QAOA evolution respect the global $\Ztwo \times \Ztwo$ symmetry and $\tau_{h/v}$ always anticommutes with $\Wilson{v/h}$.

\section{Conclusions}\label{sec:conclusion}

In this article, we presented a method to study the ground-state properties of a two-dimensional $\Ztwo$ lattice gauge theory using the quantum approximate optimization algorithm. With this method, we are able to get good quality variational approximations while keeping circuits with a small depth.
Hence, this allows to prepare the target state with a number of standard single-qubit rotations and CNOT gates comparable with the realistic expectations for near-term quantum technologies.

We focused on the minimal resources needed for an accurate description of the ground state in a quantum circuit setup, to show that interesting physics can be indeed observed despite the small size.
In particular, we showed that both the behavior of Wilson loops and the entanglement entropy clearly distinguish the trivial and the topological phase, and they characterize the confinement/deconfinement transition as well.
To reliably find good approximations of the ground state, we proposed a two-step protocol for QAOA, which produces regular optimal schedules that can be successfully transferred to larger sizes.
In this respect, the two-step protocol outperforms a resource-costly global minimum search, as well as other local optimizations strategies that are prone to remaining trapped in ``bad'' local minima.

However, the role of the noise brought by measurements and gates has been neglected, even though it will inevitably appear in a realistic implementation of our proposal.
In general, the effect of noise on VQAs is still an important open question~\cite{Cerezo_NatRev2021} and thus the study of the robustness of our proposal is worthy of further investigation.
In the worst-case scenario, where noise prevents an accurate reconstruction of the cost function $E_\Ptrot (\bgamma, \bbeta)$ for optimization purposes, one might still use a ``simulated'' QAOA to infer good variational parameters to be provided to the actual quantum circuit, which will be used mainly for measuring physical properties.

We emphasize that the QAOA technique we propose can be easily combined to extend several proposals for the study of 2D LGTs through digital quantum simulations \cite{tagliacozzo2013,tagliacozzo_Natcom2013,muschik2017u,Zohar_PRL2017,Zohar_PRA2017,Bender_NJP2018,lamm,Paulson_PRX2021,Halimeh_PRX2021}. 
Digital quantum simulations of LGTs on small systems have already been implemented in trapped ion experiments \cite{Martinez_Nature2016, Kokail_Nature2019} and superconducting qubit platforms \cite{Klco_PRA2018, Klco_PRD2020, Atas_2021, Ciavarella_PRD2021}. These experiments inspired several theoretical studies aimed at investigating the dynamics of the most important LGT excitations \cite{Pichler_PRX2016,Surace_PRX2020,Bender2020,Surace_NJP2021,Rigobello_2021}.
Our results provide a tool to efficiently initialize the ground states of LGTs, which, in turn, make it possible to engineer in a controlled way several of the excited states studied to explore the dynamical and topological properties of LGTs, including, for example, flux excitations and mesons.
The system we considered can be regarded as a surface code perturbed by onsite interactions that provide a kinetic energy to its plaquette excitations \cite{Vidal2009,Wootton2011}. Hence, the study of its dynamics delivers information on the resilience of topological quantum memories in which anyons acquire a non-trivial dispersion.
Furthermore, the topological order of the $\mathbb{Z}_2$ LGT is the same of the most common topological quantum spin liquids, and our QAOA approach can be extended, for instance, to the study of quantum dimer models based on plaquette interactions, such as the Rokhsar Kivelson model \cite{Rokhsar1988}, which displays this kind of topological phases and transitions on suitable lattices \cite{Fradkin_book}.

More in general, our variational quantum optimization successfully enables to explore the properties of Hamiltonians with non-trivial four-body interactions, which represent not only an essential element for designing topological phases but also a useful tool for encoding classical optimization problems~\cite{Lechner_SciAdv2015,Lechner_IEEE2020,Ender_arXiv2021,Dlaska_PRL2021}. Such interactions are compatible with the native geometry and qubit gate connectivity of several recently developed quantum computation platforms, encompassing both two-dimensional superconducting architectures, such as the Google Sycamore array~\cite{Arute_Nature2019,Satzinger_Science2021}, and programmable arrays of Rydberg atoms \cite{Ebadi_Nature2021,Scholl_Nature2021,Ebadi_arXiv2022,Bluvstein_arXiv2021}.
In these systems, no additional overhead would be needed to map logical into physical qubits and measurements would give direct information on the addressed models, as in the case of the $\Ztwo$ LGT.

In conclusion, the combination of QAOA, initialization of the excitations and digital quantum simulation of their time evolution opens the path to study many aspects of the dynamics of the confined and deconfined phases in LGTs as well as the anyonic excitations appearing in topologically ordered phases.

\section*{Acknowledgments}
We thank F.M. Surace for insightful discussions and S. Pradhan for sharing useful comments and exact diagonalization data about the ground states of the LGT.

M.W. and M.B are supported by the Villum Foundation (Research Grant No. 25310). This project has received funding from the European Union’s Horizon 2020 research and innovation program under the Marie Sklodowska-Curie grant agreement No. 847523 “INTERACTIONS.” 
E.E. is partially supported by INFN through the project “QUANTUM” and QuantERA ERA-NET Co-fund in Quantum Technologies (GA No. 731473), project QuantHEP. L.L. has received funding from the 2020/21 ERASMUS+ Mobility for Traineeship program.
G.E.S. was partly supported by EU Horizon 2020 under ERC-ULTRADISS, Grant Agreement No. 834402, 
and his research has been conducted within the framework of the Trieste Institute for Theoretical Quantum Technologies (TQT).
GM acknowledges support from Austrian Science Fund through the SFB BeyondC Project No. F7108-N38.

\appendix

\section{Realization of the plaquette rotation in the $3 \times 3$ torus}\label{app:3x3}

When considering a system with periodic boundary conditions and an odd number of rows and columns, some further detail must be considered when implementing the two-plaquette rotation.

\begin{figure}
    \centering
    \includegraphics[width=8.5cm]{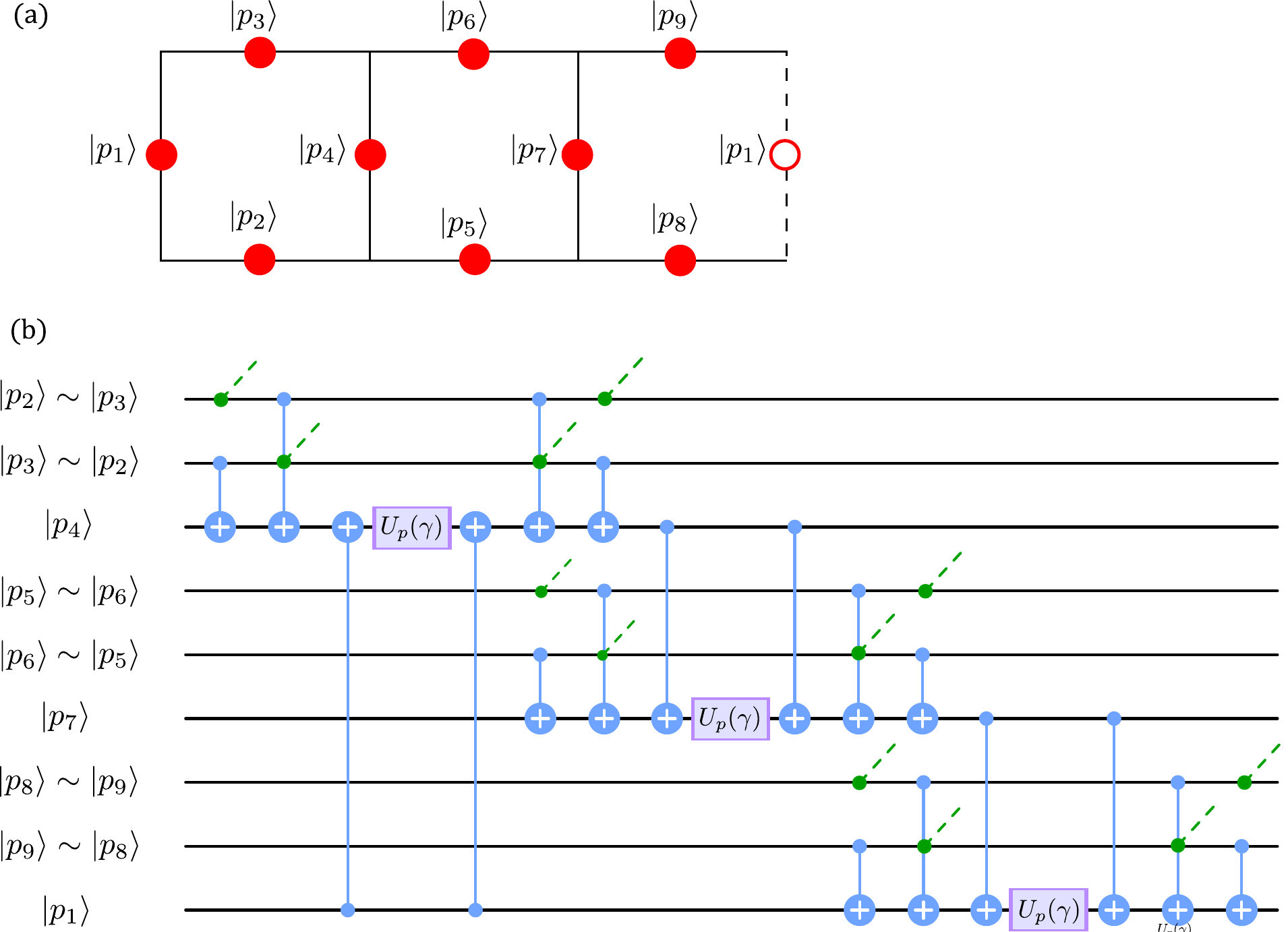}
    \caption{(a) A stripe of the $3 \times 3$ lattice with periodic boundary conditions (b) Quantum circuit to implement the plaquette operator $\nep^{i\gamma \calB}$ on all three plaquettes. The labelling of the qubit lines emphasizes that all the boundary qubits are shared with the plaquette stripes above and below. The partially depicted green CNOTS are related to the simultaneous implementation of the same algorithm on the neighboring stripes: qubits 2,3,5,6,8,9 act as controls also for the circuit in neighboring plaquettes.}
    \label{fig:threeplaquettes}
\end{figure}

In Fig. \ref{fig:threeplaquettes} we depict a circuit which generalizes the one showed in Sec. \ref{sec:alg} to the case of the $3 \times 3$ system considered throughout most of the paper. To this purpose we consider a stripe of three plaquettes, as depicted in panel (a). The operator $\nep^{i\beta \calB_p}$ is applied to all three plaquette through the algorithm displayed in Fig. \ref{fig:threeplaquettes} (b), thus with a circuit of depth 17.

When considering larger systems with periodic boundaries and an odd number of rows and columns, a suitable combination of the schemes presented in Fig. \ref{fig:twoplaquettes} and \ref{fig:threeplaquettes} allows us to perform each QAOA step with a circuit of depth 18 involving only CNOTS between neighboring qubits and single-qubit rotations.

\section{Details on the QAOA implementation}\label{app:QAOA}
In this appendix we discuss additional details regarding our two-step implementation of QAOA for the problem under investigation.
In particular, we focus on a benchmark of our heuristic approach against a global minimum search, which, remarkably, yields similar-quality results for both phases, in terms of ground state fidelity, offering a  good numerical validation of our scheme.

In addition, we comment on the transferability of the optimal schedules, obtained by either two-step or basin hopping, to larger system sizes, a strategy that could provide an educated guess to lower the computational cost for a new optimization.

Finally, we observe some patterns for optimal QAOA variational parameters obtained with the two-step scheme, in particular their smoothness as a function of $m=1\cdots \Ptrot$, similarly to other results for different QAOA applications~\cite{Zhou_PRX2020,Mbeng_arx19, Pagano_PNAS2020}.

\subsection{Global optimization vs two-step scheme}\label{app:globalvslocal}
In order to prove the effectiveness of the two-step optimization protocol, we compare it with a global minimum search, based on the basin hopping method~\cite{Wales_basinhopping} from the scipy.optimize Python library.
In the latter algorithm, we allow up to 500 local minimizations, each of them initialized in the proximity of a previously found local minimum. The parameter space is explored with an effective temperature chosen to allow jumps between typical low energy minima. To reliably find the absolute minimum we run the basin hopping optimization  100 times and take the best result.

\begin{figure}
    \centering
    \includegraphics[width=8.5cm]{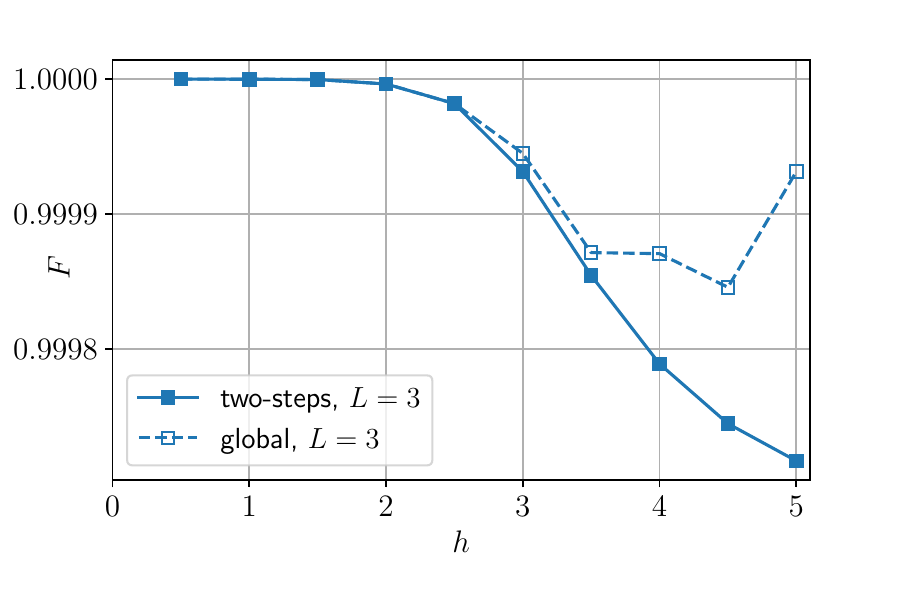}
    \caption{ Comparison of the accuracy with respect to the magnetic coupling $h$ between global optimization (empty squares) and the two-step approach (full squares) described in the text. 
    The data refer to $\Ptrot=6$ and $\ket{\psi_0}=\ket{\Omega_E}$.}
    \label{fig:Fidelity-globalvsdt}
\end{figure}
Figure~\ref{fig:Fidelity-globalvsdt} shows a comparison between the fidelity obtained with the global and the two-step optimizations, as a function of $h$ for fixed $\mathrm{P}=6$. The initial state is $\ket{\psi_0} = \ket{\Omega_E}$.
For $h<3$ the two-step approach can match the global optimization performance and it yields the same results, while for $h \geq 3$ it finds a sub-optimal local minimum.
However, we stress that, even in this case, the final fidelity is almost one: the difference in the accuracy between the two methods is much lower than a realistic experimental resolution.
Moreover, the two-step protocol has the clear advantage of requiring only a single local optimization --- on top of a modest computational overhead for the one-dimensional optimal $\Delta t$ grid search --- to be compared with $500 \times 100$ local optimizations for the basin hopping method. 
Consequently, the two-step heuristics certainly requires drastically fewer function evaluations and it is, therefore, a better candidate to be implemented on a realistic quantum device and also much faster to simulate on a classical computer.

Regarding the transferability of the optimal schedules to larger system sizes, we use the optimal angles obtained for $L=3$, either with basin hopping or with a two-step optimization, as initial guess for local optimizations of larger system sizes $L=4,5$. 
Specifically, for each value of $h$, we compare the best fidelity out of 10 BFGS local search runs, each of them initialized with the optimal $2\mathrm{P}$ parameters previously found for $L=3$, plus a small noise to facilitate the exploration of the energy landscape. 

\begin{figure}
    \centering
    \includegraphics[width=8.5cm]{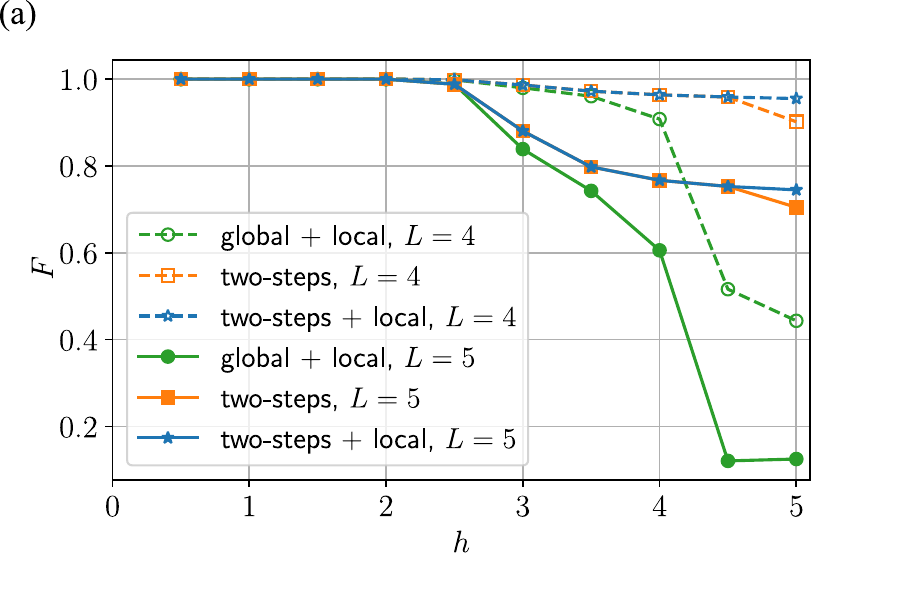}
    \includegraphics[width=8.5cm]{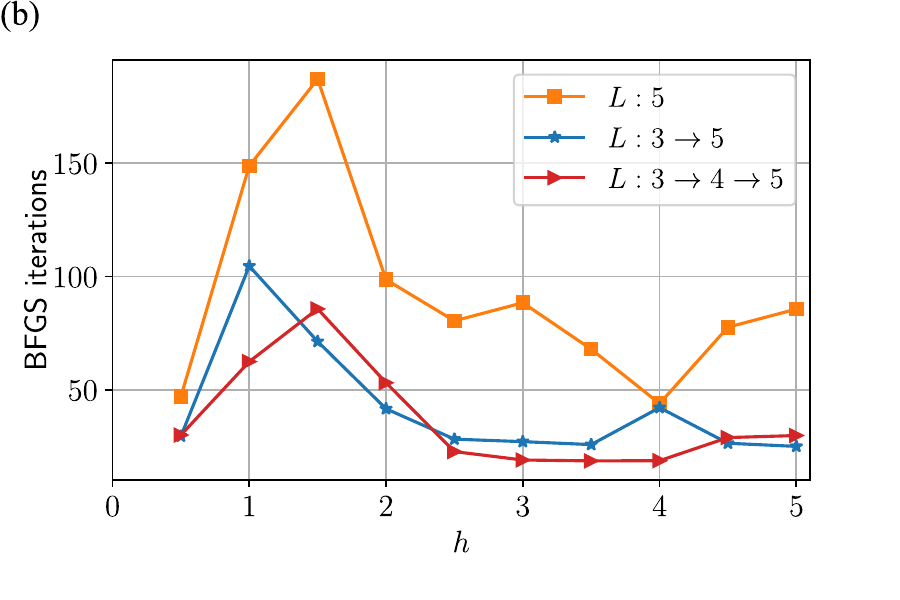} 
    \caption{(a) Comparison of the two-step vs global schedules for $L=3$ as an educated guess for optimization on larger system sizes $L=4,5$. The data refer to the best out of 10 local BFGS minimizations, run by starting close to the optimal two-step schedule (blue stars) or optimal global-schedule (green circles). The plot also shows the fidelity of optimal two-step schedule (orange squares) for $L=4,5$.
    (b) Number of iterations required for the convergence of the final BFGS optimization; comparison between transferring the schedule from $L=3$ to $L=5$ (blue stars), transferring from $L=4$ to $L=5$ (red triangles), and two-steps optimization directly on $L=5$ (orange squates).
    All data refer to $\Ptrot=6$ and $\ket{\psi_0}=\ket{\Omega_E}$.}
    \label{fig:Fidelity-globalvsdt-initialization}
\end{figure}

Our results are reported in Fig.~\ref{fig:Fidelity-globalvsdt-initialization}(a), where we compare local minimizations starting from the $L=3$ two-step optimal schedules (star symbols) local searches starting from to the $L=3$ global minimum (circles), and the two-step process applied directly on the lerger system (squares).
We find that the optimal angles returned by the two-step algorithm provide a better guess for larger systems,  resulting in higher fidelity than a local search initialized with the global minimum for $L=3$. This may be linked to the existence of some patterns in the optimal parameters found with the two-step scheme, in particular their smoothness as a function of $m=1\cdots \Ptrot$, as summarized in Appendix~\ref{app:smooth}.
The performance of the two-step optimization applied directly on the target system $L=4$ or $L=5$ is instead comparable with the transfer of the schedule from $L=3$, although the latter is slightly better at large magnetic fields.

However, once the optimal schedule for a given system size is known, it is convenient to leverage on that result to initialize the QAOA search for larger systems, instead of running a new two-step optimization from scratch.
In fact, although the performances in terms of final fidelity are similar, the schedule transfer requires less iterations than the two-steps optimization. 
This is shown in Fig.~\ref{fig:Fidelity-globalvsdt-initialization}(b), where we compare the number of BFGS iterations required in the final local minimum search on a system with $L=5$ for different optimization strategies: schedule transferring from $L=3$ to $L=5$, from $L=4$ to $L=5$, and the two-steps protocol directly on $L=5$.
The latter requires in general a larger number of iterations and its overall cost must be added to the resources required for the optimization of the time step $\Delta t$.

\subsection{Smooth schedules}\label{app:smooth}
\begin{figure*}
    \centering
    \includegraphics[width=16cm]{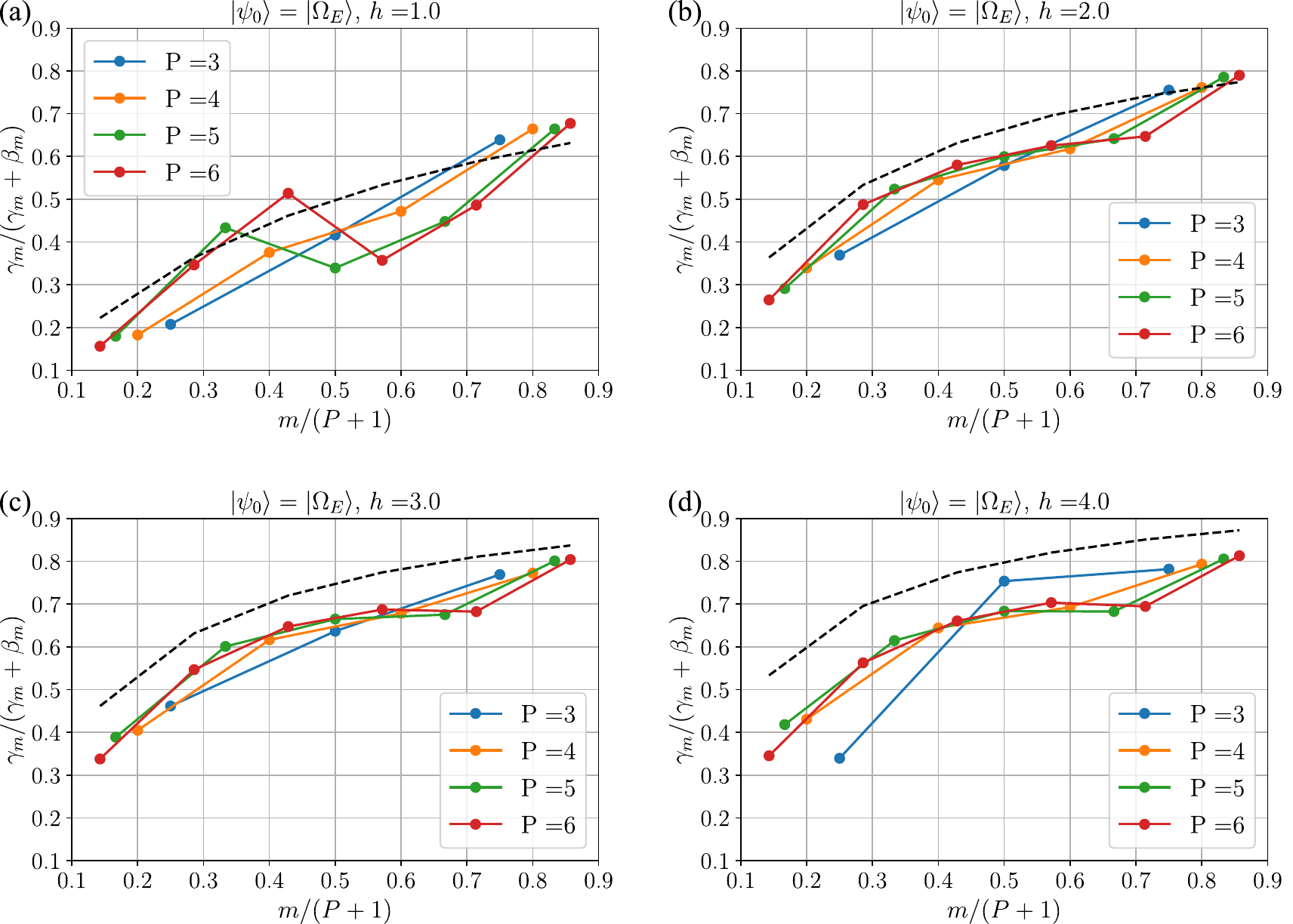}
    \caption{Variational parameters associated to the best result out of 10 local minimum searches, following the two-step optimization procedure.
    In each panel, the black dashed line corresponds, through Eq.~\eqref{eq:dQA_s_m_Delta_t}, to the linear annealing schedule defined by Eq.\eqref{eq:dQA_el}, with the corresponding value of $h$ and $\Ptrot=6$ steps.
    The quantum circuit is initially prepared in the electric ground state $\ket{\Omega_E}$.
    }
    \label{fig:param_dt-opt}
\end{figure*}
In the $\Ztwo$ LGT we studied in this paper, we found that the two-step optimization scheme produces smooth protocols for the optimal variational parameters more easily than other heuristic methods present in the literature, such as the standard application of iterative schemes based on parameter interpolation or Fourier component optimization~\cite{Mbeng_arx19,Zhou_PRX2020}.
Moreover, it provides a minimum $(\bgamma^*,\bbeta^*)$ for a chosen circuit depth $\Ptrot$ without requiring the solution for shallower circuits with $\Ptrot' < \Ptrot$, contrarily to both the iterative methods just mentioned. 

The presence of regular patterns in the optimal parameters suggests a 
comparison with a digitized quantum annealing scheme, such as the ones adopted to initialize the two-step QAOA.
In Sec.~\ref{sec:results_ts} we defined two possible annealing protocols, depending on the choice of the initial state.
If $\ket{\psi_0}=\ket{\Omega_E}$, we construct the time dependent Hamiltonian
\begin{eqnarray}\label{eq:Hann_el}
\Ham(t)=\Ham_E + h \frac{t}{\tau} \Ham_B \,,
\end{eqnarray}
while if $\ket{\psi_0}=\ket{\Omega_B}$ we use
\begin{eqnarray}\label{eq:Hann_magn}
\Ham(t)=h\left(\frac{t}{h\tau}\Ham_E +  \Ham_B \right) \,.
\end{eqnarray}
In both cases, $t\in [0,\tau]$ and at the end of the protocol $\Ham(t=\tau)=\Htarg$.
The corresponding parameters $\gamma_m$ and $\beta_m$ of a digitized quantum annealing are reported in Eq.~\eqref{eq:dQA_el} and Eq.~\eqref{eq:dQA_magn}, respectively.

For a graphical representation of smooth optimal two-step schedules and a direct comparison with digitized Quantum Annealing, it is useful to consider the following more general protocol, as customary in Adiabatic Quantum Computation~\cite{Albash_RMP2018}:
\begin{equation}
    \Ham(s) = (1-s)\Ham_E + s\Ham_B \ ,
\end{equation}
where $s(t)\in[0,1]$ is a monotone time-dependent parameter that interpolates between $\Ham_E$ and $\Ham_B$.
With this notation, we can identify Eq.~\eqref{eq:Hann_el} with a process starting from $s(0)=0$ and ending in $s(\tau)=s_\text{\scriptsize f}$, with $s_\text{\scriptsize f}=\frac{h}{h+1}$;
Eq.~\eqref{eq:Hann_magn}, instead, corresponds to a process with $s(0)=1$, ending again in $s(\tau)=s_\text{\scriptsize f}$ (both identifications are valid modulo an overall multiplicative factor).

A digitized Quantum Annealing process~\cite{Mbeng_arx19} consists in choosing a discretization of the time interval $[0,\tau]$ into $\Ptrot$ small time steps $\Delta t_m$, such that $\sum_m \Delta t_m= \tau$. 
Correspondingly, the continuous schedule $s(t)$ is discretized into a sequence of short-time evolutions generated by $\Ham(s_m)$, where
\begin{equation} \label{eq:s_m}
s_m=s_0+(s_\text{\scriptsize f}-s_0)\, \frac{m}{\Ptrot} \;,   
\end{equation}
with $m=1\cdots \Ptrot$.
The resulting expression can be further simplified with a first order Trotter split-up, neglecting quadratic terms in $\Delta t_m$.
Thus, the final state is written as the variational {\em Ansatz} in Eq.~\eqref{eq:QAOA_state}, with \emph{fixed} parameters given by
\begin{equation} \label{eq:dQA_s_m_Delta_t}
    \left\lbrace 
    \begin{split}
        s_m = \frac{\gamma_m}{\gamma_m+\beta_m} \ ,\\
        \Delta t_m = \gamma_m + \beta_m \ .
    \end{split}
    \right.
\end{equation}

Once we have found optimal smooth QAOA parameters $\bgamma^\star,\; \bbeta^\star$ with our two-step QAOA scheme discussed in Sec.~\ref{sec:results_ts}, we can extract the corresponding digitized schedule $s^*_m$ and compare it with the linear digitized quantum annealing protocol $s_m^\dQA$ that we used as an educated guess for the local minimization.

As examples of typical smooth QAOA optimal parameters, we report in Fig.~\ref{fig:param_dt-opt}(a)-(d) the schedules $s^*_m$ corresponding to four different values of the coupling $h$, both below and above the ``topological transition'', with $\Ptrot \ge 3$ and initial state $\ket{\psi_0}=\ket{\Omega_E}$.
The dashed black lines correspond to the linear annealing schedule of Eq.\eqref{eq:dQA_el} we used as starting point for the local minimizations, with $\Ptrot=6$.
In all four cases, it appears evident that as $\Ptrot$ increases, the parameters gradually approach a smooth continuous behavior, with the possible exception
of a single localized irregularity, which seems to appear in Fig.~\ref{fig:param_dt-opt}(a) for $\Ptrot=5,6$.
This is not surprising, however, since we are preparing a state very close to the initial one. Thus, a large value of $\Ptrot$ could ``overfit'' the target state and many different parameters choices, usually non-smooth, could yield similar accuracy. 
A comparison with Fig.~\ref{fig:QAOA_opt} for the case $h=1$, clearly shows a degradation of performance (almost-flat curve) of the infidelity vs $\Ptrot$, exactly for $\Ptrot=5,6$: this irregularity can thus be interpreted as a local lower-quality minimum or a saturation of the numerical precision of the algorithm.

For larger values of the coupling $h$, instead, we observe a clear continuity in the optimal schedule $s_m^*$, as we change both $\Ptrot$ and $h$.
This leads to the interesting consequence that the optimal schedule for a given $\Htarg$ and circuit depth $\Ptrot$ could be used as a seed to initialize the optimization for different values of $h$, requiring only a small fine-tuning of the parameters to adapt the schedule to the new target ground state.

\begin{figure*}
    \centering
    \includegraphics[width=16cm]{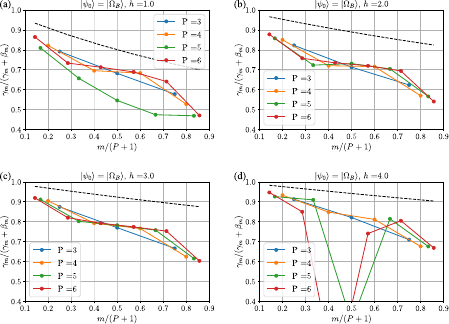}
    \caption{Variational parameters associated to the best result out of 10 local minimum searches, following the two-step optimization procedure.
    In each panel, the black dashed line corresponds, through Eq.~\eqref{eq:dQA_s_m_Delta_t}, to the linear annealing schedule defined by Eqs.\eqref{eq:dQA_magn}, with the corresponding value of $h$, and $\Ptrot=6$ steps.
    The quantum circuit is initially prepared in the magnetic ground state $\ket{\Omega_B}$.
    }
    \label{fig:param_dt-opt_reverse}
\end{figure*}
A similarly smooth pattern is observed when we initialize the system in the magnetic ground state $\ket{\Omega_B}$, as reported in Fig.~\ref{fig:param_dt-opt_reverse}(a)-(d).
The dashed black lines correspond here to the schedules $s_m^\dQA$ extracted from Eq.\eqref{eq:dQA_magn}, with $\Ptrot=6$.
The main difference is that the smoothness now is more easily lost when targeting the deconfined phase, see panel(d), which is closer to the initial state. Similar comments on this irregularity apply as for the previous case, by comparing with Fig.~\ref{fig:QAOA_opt_reverse}(a).
On a side note, we notice that the evident irregularity in panel(d) for $\Ptrot=5,6$ involves a single point with a numerical value smaller than 0.4: this is not a significant feature, and it could easily be eliminated by an appropriate smoothing procedure with a likely improve in performance. 

Unsurprisingly, the two-step optimization might get trapped in a (high-quality) local minimum even when we target the opposite phase: this is seen, e.g., for the outlier set of $h=1$ and $\Ptrot=5$ in Fig.~\ref{fig:param_dt-opt_reverse}(a), which might be associated to a suboptimal minimum.
This observation is once again consistent with the corresponding data in Fig.~\ref{fig:QAOA_opt_reverse}(a), where the curve for $h=1$ shows 
a small spike in correspondence to $\Ptrot=5$.

Regarding the comparison with the linear dQA protocol $s_m^\dQA$ (dashed black lines), in both Figs.~\ref{fig:param_dt-opt} and \ref{fig:param_dt-opt_reverse}, the overall monotonicity of optimized $s^*_m$ is the same of the original schedule, i.e., an increasing function of $m$ when $\ket{\psi_0}=\ket{\Omega_E}$, and a decreasing function when $\ket{\psi_0}=\ket{\Omega_B}$.
However, when targeting states in a phase that differs from the initial one, the optimal schedule deviates more and more from the original {\em Ansatz}, highlighting the importance of the local optimization of the parameters.

\subsection{Energy landscape}\label{app:energy}
In the $\Ztwo$ LGT model, the energy landscape associated to the QAOA {\em Ansatz} is characterized by the presence of many local minima, as discussed in Sec.~\ref{ssec:energy_landscape}.
This makes the employment of a clever optimization strategy, such as the two-step protocol or schedule transferability, a necessity to target reliably low energy minima.
However, for general variational problems, it might happen that it exists a deep minimum in the energy landscape, associated to a state with small or no overlap with the target one.
This is not the case for the problem under investigation, where there is a clear correlation between the fidelity and the variational energy for the minima in the energy landscape, see Fig.\ref{fig:randominit} in the main text.

\begin{figure}
    \centering
    \includegraphics[width=8.5cm]{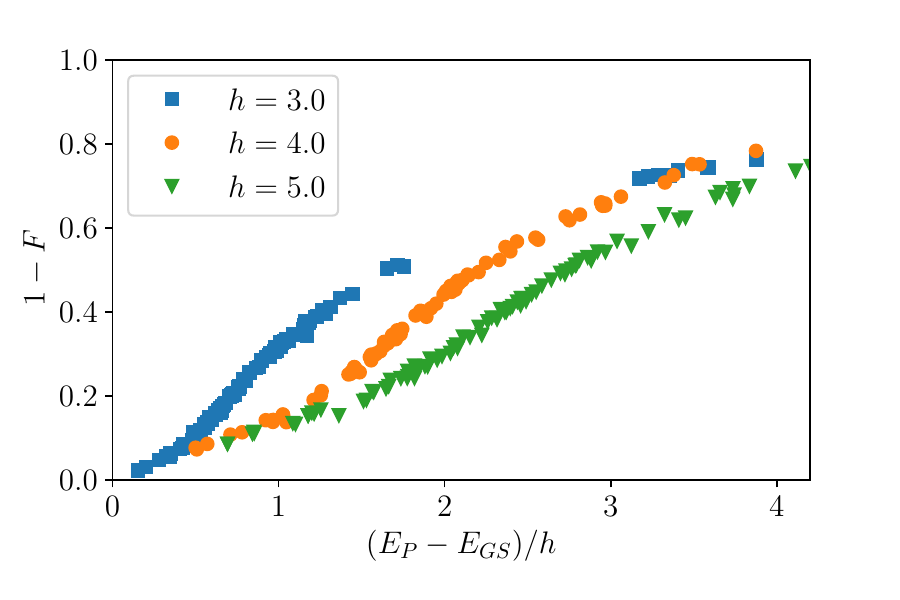}
    \includegraphics[width=8.5cm]{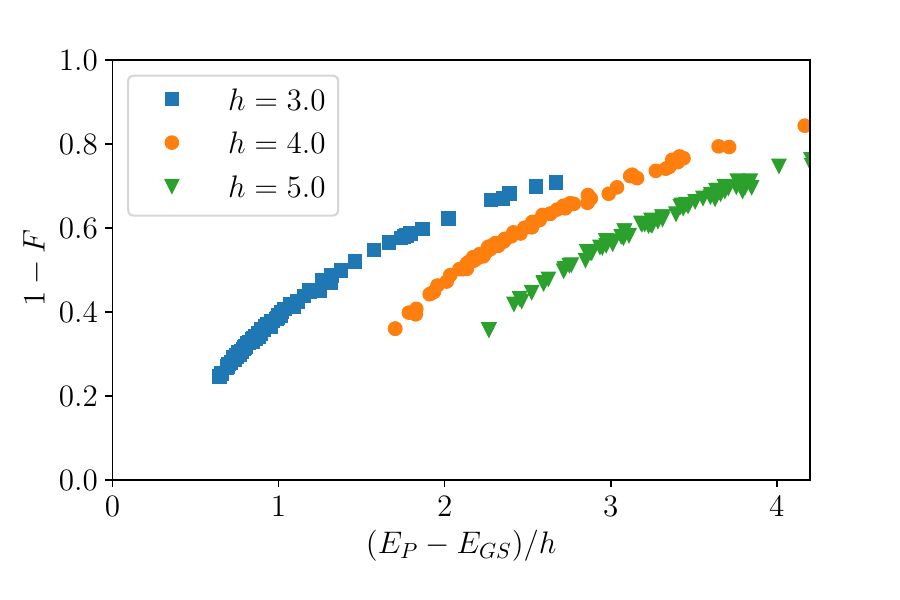}
    \caption{Infidelity vs residual energy for 100 minima in the energy landscape found with random initialization of local BFGS searches, with $\ket{\psi_0}=\ket{\Omega_E}$ and $\Ptrot=6$.
    Panel a. corresponds to the lattice with $4 \times 4$ plaquettes, panel b. to $5 \times 5$ plaquettes. We only show the data that falls in the interval $(E_P - E_{GS})/h \in [0, 4.2]$.}
    \label{fig:random_app}
\end{figure}

Here, we show that this correlation holds also for the larger systems considered in this paper, $L=4$ and $L=5$, corresponding to 32 and 50 qubits respectively.
We repeat the analysis of Sec.~\ref{ssec:energy_landscape}: focusing on $\ket{\psi_0}=\ket{\Omega_E}$, we perform 100 QAOA runs with random initial parameters, targeting states in the deconfined phase $h\ge 3$.
In Fig.~\ref{fig:random_app} we plot the infidelity $1-F_\Ptrot$ vs the residual energy of the minima found with this procedure, for both $L=4$ (a. panel) and $L=5$ (b. panel) and a circuit depth of $\Ptrot=6$.
Interestingly, it appears that increasing the system size leads to a sharper correlation between energy and fidelity, compared to the data presented in Fig.~\ref{fig:randominit}.

%

\end{document}